\documentclass[useAMS,usenatbib]{mn2e}
\usepackage{epsf}
\usepackage{natbib}
\usepackage{epsfig}
\usepackage{subfigure}

\newcommand{\gimic}        {\textsc{gimic}}

\newcommand{\gadget}      {\textsc{gadget-3}}
\newcommand{\subfind}    {\textsc{Subfind}}

\title[{Structure and kinematics of stellar haloes}]{Global structure and kinematics of stellar haloes in cosmological hydrodynamic simulations}

\author[I.~G.~McCarthy et~al.]{I. G. McCarthy$^{1,2}$\thanks{E-mail:
mccarthy@ast.cam.ac.uk},  A.~S.~Font$^{2}$, R.~A.~Crain$^{3,4}$, A.~J.~Deason$^{2}$, J.~Schaye$^{4}$, \newauthor T.~Theuns$^{5,6}$ \\
$^{1}$Kavli Institute for Cosmology, University of Cambridge, Madingley Road, Cambridge, CB3 OHA\\
$^{2}$Institute of Astronomy, University of Cambridge, Madingley Road, Cambridge, CB3 0HA\\
$^{3}$Centre for Astrophysics \& Supercomputing, Swinburne University of Technology, Hawthorn, Victoria 3122, Australia\\
$^{4}$Leiden Observatory, Leiden University, P. O. Box 9513, 2300 RA Leiden, the Netherlands\\
$^{5}$Institute of Computational Cosmology, Department of Physics, University of Durham, Science Laboratories, South Road, Durham DH1 3LE\\
$^{6}$Department of Physics, University of Antwerp, Campus Groenenborger, Groenenborgerlaan 171, B-2020 Antwerp, Belgium
}

\begin{document}

\date{Accepted ... Received ...}

\pagerange{\pageref{firstpage}--\pageref{lastpage}} \pubyear{2008}

\maketitle

\label{firstpage}

\begin{abstract}
We use the Galaxies-Intergalactic Medium Interaction Calculation (GIMIC) suite of cosmological hydrodynamical simulations to study the global structure and kinematics of stellar spheroids of Milky Way mass disc galaxies.  Font et al.\ have recently demonstrated that these simulations are able to successfully reproduce the satellite luminosity functions and the metallicity and surface brightness profiles of the spheroids of the Milky Way and M31.  A key to the success of the simulations is a significant contribution to the spheroid from stars that formed {\it in situ}.  While the outer halo is dominated by accreted stars, stars formed in the main progenitor of the galaxy dominate at $r \la 30$ kpc.
In the present study we show that this component was primarily formed in a proto-disc at high redshift and was subsequently liberated from the disc by dynamical heating associated with mass accretion.  As a consequence of its origin, the in situ component of the spheroid has different kinematics (namely net prograde rotation with respect to the disc) than that of the spheroid component built from the disruption of satellites.  In addition, the in situ component has a flattened distribution, that is due in part to its rotation.  We make comparisons with measurements of the shape and kinematics of local galaxies, including the Milky Way and M31, and stacked observations of more distant galaxies.  We find that the simulated disc galaxies have spheroids of the correct shape (oblate with a median axis ratio of $\sim0.6$ at radii of $\la 30$ kpc, but note there is significant system-to-system scatter in this quantity) and that the kinematics show evidence for two components (due to in situ vs.\ accreted), as observed.  Our findings therefore add considerable weight to the importance of dissipative processes in the formation of stellar haloes and to the notion of a `dual stellar halo'.
\end{abstract}

\begin{keywords}
Galaxy: evolution  ---  Galaxy: formation ---  Galaxy: halo --- galaxies: evolution  ---  galaxies: formation ---  galaxies: haloes
\end{keywords}

\section{Introduction}
\label{sec:intro}

It has long been recognised that the extended, diffuse stellar haloes of galaxies contain a potential treasure trove of information about the formation and evolution of galaxies (e.g., \citealt{eggen62,searle78}).  Observationally, we are entering a `golden age' for studies of stellar haloes, with the exploitation of large surveys such as the SDSS / Sloan Extension for Galactic Understanding
and Exploration\footnote{http://www.sdss3.org/} (SEGUE, \citealt{yanny09}), RAVE\footnote{http://www.rave-survey.aip.de/rave/}, SkyMapper\footnote{http://www.mso.anu.edu.au/skymapper/}, Pan-Starrs\footnote{http://pan-starrs.ifa.hawaii.edu/public/} and Gaia\footnote{http://gaia.esa.int/} that are yielding (or will yield) exquisite structural and kinematic measurements of the Milky Way's stellar halo and its substructures.  At the same time, dedicated surveys of nearby galaxies, such as M31 (e.g., \citealt{guhathakurta05,irwin05,ibata07}), NGC 891 \citep{ibata09,mouhcine10} and other nearby galaxies (e.g., in the GHOSTS survey\footnote{http://www.stsci.edu/~djrs/ghosts/}), provide us with a valuable external perspective and a point of comparison for the Milky Way (i.e., how typical is the Milky Way?).

Over the past few years there have also been significant advances in the modelling of stellar haloes of galaxies with masses similar to that of the Milky Way.  The majority of these studies have relied on so-called `hybrid' models, which couple simple recipes for assigning stellar mass to infalling satellites in dissipationless (i.e., dark matter only) cosmological simulations (e.g., \citealt{bullock05,diemand05,font06a,delucia08,cooper10}).  In terms of assigning stellar mass to the infalling satellites, the models are generally tuned to reproduce the observed luminosity function of satellites around the Milky Way.  The stellar mass is distributed by tagging centrally-located dark matter particles and tracking the DM particles using their IDs.  In general, the models have been largely successful at reproducing the observed outer stellar halo, as well as the observed demographics of stellar streams.  

However, it has also become clear that such hybrid models have a number of problems.  For example, \citet{font06b}, \citet{delucia08}, and \citet{cooper10} have all shown that they fail to produce significant metallicity gradients in the stellar spheroids, whereas observations of the Milky Way (e.g., \citealt{carollo07,carollo10,dejong10}, but see \citealt{sesar11}) and M31 (\citealt{kalirai06,koch08,gilbert09}) show evidence for large-scale gradients\footnote{There is a degree of confusion in the literature over the term ``metallicity gradient''. Detailed observations of main-sequence turn-off stars (e.g., \citealt{carollo07,carollo10,dejong10} and blue horizontal-branch stars (e.g., \citealt{beers11}) in the Milky Way show evidence for two components (one relatively metal rich and the other relatively metal poor) in the metallicity distribution function (MDF) that overlap spatially in the inner regions of the halo. In the outer regions, beyond $r \sim 20$ kpc or so, however, the metal-poor component is dominant and there is no further alteration of the MDF as one progresses further into the outer halo (i.e., no gradient). The transition from metal-rich component dominance to metal-poor component dominance would suggest a gradient if these populations are not separated explicitly (e.g., as in the case of observations of M31), because of the lessening degree of dominance between the metal-rich component, relative to the metal-poor component, as one moves outward. It is in this sense that we use the term gradient. \citet{font11} demonstrated that the simulated galaxies in \gimic\ exhibit exactly this behavior; i.e., that there are two populations (in situ and accreted) with different characteristic metallicities that overlap spatially, and it is the change in the mass fraction of these populations with radius that gives rise to an overall gradient out to radii of 30 kpc (see their Fig. 8). Beyond this radius, the accreted component dominates by mass, thus there is no further decline in the mean metallicity.}. Furthermore, observations of the Milky Way (e.g., \citealt{chiba01}) and M31 (e.g., \citealt{pritchet94}), as well as stacked observations of distant galaxies \citep{zibetti04}, show that the spheroid has a highly flattened (oblate) distribution, whereas the stellar haloes in hybrid simulations usually have a {\it prolate} configuration like the dark matter \citep{cooper10}.  Observations of the spheroid in the Solar neighbourhood show kinematic and chemodynamic evidence for two distinct components to the spheroid (the `dual halo' e.g., \citealt{carollo07,carollo10,nissen10}) that may be difficult to reconcile with purely dissipationless hybrid methods.  Finally, hybrid models appear to overproduce the degree of substructure seen in the stellar halo of the Milky Way (e.g., \citealt{helmi11,xue11}).

Hybrid models implicitly assume that the spheroid is built entirely of the tidal disruption of infalling satellites, they do not include/allow for dissipational processes such as in situ star formation (i.e., star formation in the spheroid itself, or star formation in a disc which is subsequently heated to become a spheroid).  Studies of the formation of stellar haloes in cosmological hydrodynamic simulations, however, routinely show that this channel for producing stellar spheroids is significant (e.g., \citealt{abadi06,zolotov09,zolotov10,oser10,scannapieco11,font11}, hereafter Paper I).  An important caveat to bear in mind, is that many cosmological simulations suffer from a large degree of overcooling, and the importance of the in situ component could thus be overestimated in such simulations.  However, as we showed in Paper I and in \citet{crain10} (hereafter C10), the \gimic\ suite of cosmological simulations \citep{crain09} does not suffer from a large degree of overcooling on the scale of Milky Way mass (and lower) systems, due to efficient (but energetically-feasible) supernova (SN) feedback.  Note that strong SN feedback is widely believed to be required to explain the generally low star formation efficiency (e.g., \citealt{croton06,bower06,schaye10}) and the `missing baryon problem' (e.g., \citealt{bregman07}; C10) of normal galaxies, as well as the enrichment of the intergalactic medium (e.g., \citealt{aguirre01,oppenheimer06,wiersma11}).
As we demonstrated in Paper I, the \gimic\ simulations yield satellite luminosity functions and stellar spheroid surface brightness distributions that are quite comparable to those observed for the Milky Way and M31, which is a direct result of efficient SN feedback.

We also demonstrated in Paper I that the \gimic\ simulations successfully reproduce the radial distribution of metals (i.e., the large-scale gradient) in the spheroid, a problem that has been a thorn in the side of hybrid methods for some time.  This begs the question of whether or not a significant in situ component can also alleviate the problems that hybrid methods have in reproducing the shape of the spheroid, the evidence for a `dual halo' from kinematics, and also the degree of substructure.  

In the present study, we focus on the global structure and kinematics of spheroids and leave the study of substructure for a future work.  In particular, in the present study we first examine in some detail the nature of the in situ component in the \gimic\ simulations and show that its origins are deeply rooted in the dynamical heating of the proto-disc at high redshift.  The implication of this finding is that the in situ component of the spheroid is expected to have different kinematic characteristics (namely rotation and perhaps also a different velocity anisotropy profile) than that of the spheroid component built from the disruption of satellites.  In addition, one expects a more flattened distribution for this in situ component, due to its net rotation.  We make comparisons with measurements of the shape and kinematics of the Milky Way, M31, and more distant galaxies.  The upshot of this comparison is that the simulated disc galaxies have spheroids of the correct shape (oblate with a median axis ratio of $\approx 0.6$ at $r \la 30$ kpc) and that the kinematics show evidence for two components (due to in situ vs.\ accreted), as observed.

The present paper is organised as follows.  In Section \ref{sec:sims} we describe the simulations and the selection of the sample of Milky Way-mass disc galaxies.  In Section 3 we examine the origin of the in situ component of the spheroid in the simulations.  In Section 4 we present our main results on the global structure and kinematics of the spheroids of the simulated galaxy population and make comparisons with observations.  Finally, in Section 5 we summarise and discuss our findings.

\section{Simulations and Sample Selection}
\label{sec:sims}

We use the Galaxies-Intergalactic Medium Interaction Calculation (\gimic) suite of cosmological hydrodynamical simulations, which were carried out by the Virgo Consortium and are described in detail by \citet{crain09}, hereafter C09 (see also C10 and Paper I).  We present only a very brief summary of the simulations here and refer the interested reader to the above papers for further details.

The suite consists of a set of hydrodynamical re-simulations of five nearly spherical regions ($\sim 20 h^{-1}$ Mpc in radius) extracted from the Millennium Simulation \citep{springel_etal_05}.  The regions were picked to have overdensities at $z=1.5$ of $(+2, +1, 0, -1, -2) \sigma$, where $\sigma$ is the root-mean-square deviation from the mean on this spatial scale.  The 5 spheres are therefore environmentally diverse in terms of the large-scale structure that is present within them.  For the purposes of the present study, however, we will only select systems with total `main halo' (i.e., the dominant subhalo in a friends-of-friends, hereafter FoF, group) masses similar to that of the Milky Way, irrespective of which \gimic\ volume it is in (i.e., irrespective of the large-scale environment).  C09 found that the properties of systems of fixed main halo mass do not depend significantly on the large-scale environment (see, e.g., Fig.\ 8 of that study).

The cosmological parameters adopted for \gimic\ are the same as those assumed in the Millennium Simulation and correspond to a $\Lambda$CDM model with $\Omega_{m} = 0.25$, $\Omega_{\Lambda} = 0.75$, $\Omega_{b} = 0.045$, $\sigma_{8} = 0.9$ (where $\sigma_{8}$ is the rms amplitude of linearly evolved mass fluctuations on a scale of $8 h^{-1}$ Mpc at $z=0$), $H_{0} = 100 h$ km  s$^{-1}$ Mpc$^{-1}$, $h = 0.73$, $n_s=1$ (where $n_s$ is the spectral index of the primordial power spectrum).  The value of $\sigma_8$ is approximately 2-sigma higher than has been inferred from the most recent CMB data \citep{komatsu09}, which will affect the abundance of Milky Way-mass systems somewhat, but should not significantly affect their individual properties.

The simulations were evolved from $z=127$ to $z=0$ using the TreePM-SPH code \gadget\ 
(last described in \citealt{springel05}), which has been modified to incorporate new baryonic physics. Radiative cooling rates for the gas are computed on an element-by-element basis by interpolating 
within pre-computed tables generated with CLOUDY \citep{ferland98} that contain 
cooling rates as a function of density, temperature, and redshift and that 
account for the presence of the cosmic microwave background and 
photoionisation from a \citet{haardt01} ionising UV/X-Ray background 
(see \citealt{wiersma09a}).  This background is switched on at $z=9$ in the 
simulation, where it quickly `reionises' the entire simulation volume.  Star formation 
is tracked in the simulations following the prescription of \citet{schaye08} which reproduces the observed Kennicutt-Schmidt law \citep{kennicutt98} by construction.  
The timed release of individual elements by both massive (Type II SNe and 
stellar winds) and intermediate mass stars (Type Ia SNe and asymptotic giant branch stars) 
is included following the prescription of \citet{wiersma09b}.  A set of 11
individual elements are followed in these simulations (H, He, C, Ca, N, O, Ne, Mg, S, Si, 
Fe), which represent all the important species for computing radiative cooling rates.

Feedback from SNe is incorporated using the kinetic wind model 
of \citet{dallavecchia08} with the initial wind velocity, $v_w$, set to $
600$ km/s and the mass-loading parameter (i.e., the ratio of the mass of gas 
given a velocity kick to that turned into newly formed star particles), 
$\eta$, set to $4$.  This corresponds to using 
approximately 80\% of the total energy available from SNe for a \citet{chabrier03} IMF, which is assumed in the simulation.  This choice of parameters yields a good match to the peak of the star formation rate history of the universe (C09; see also \citealt{schaye10}) and reproduces a number of X-ray/optical scaling relations for normal disc galaxies (C10).   As shown in Paper I, the \gimic\ simulations also produce realistic spheroidal components around $\sim L_*$ disc galaxies (see Figs.\ 3-5 of that paper, which compare the predicted surface brightness and metallicity profiles to observations).

To test for numerical convergence, the \gimic\ simulations have been run at three levels of numerical resolution: `low', `intermediate', and `high'.  The low-resolution simulations have the same mass resolution as the Millennium Simulation (which, however, contains only dark matter) while the intermediate- and high-resolution simulations have $8$ and $64$ times better mass resolution, respectively.  Only the $-2\sigma$ volume was run to $z=0$ at high resolution, owing to the computational expense of running such large volumes at this resolution.  As in Paper I, we adopt the intermediate-resolution simulations in the main analysis and reserve the high-resolution $-2\sigma$ simulation to test the numerical convergence of our results.  The intermediate-resolution runs have a dark matter particle mass $m_{\rm DM} \simeq 5.30 \times 10^{7} h^{-1}$ M$_{\odot}$ and an initial gas particle mass of $m_{\rm gas} \simeq 1.16 \times 10^{7} h^{-1}$ M$_{\odot}$, implying that it is possible to resolve satellites with total stellar masses similar to those of the classical dwarf galaxies around the Milky Way (with $M_{*} \sim 10^{9-10} M_{\odot}$) with several hundred up to $\sim1000$ particles.  The gravitational softening of the intermediate-resolution runs is $1 h^{-1}$ kpc.  As we showed in the Appendix of Paper I, the surface brightness and metallicity profiles are converged at this resolution, as is the satellite luminosity function for satellites brighter than $M_V \approx -12.5$ (note that bright satellites dominate the assembly of the accreted component of the stellar spheroid; see, e.g., \citealt{font06a,cooper10}).  In the Appendix of the present paper, we show the global shape and kinematics of the spheroid in the intermediate-resolution runs agrees well with that of the high-resolution run.

Note that for all these simulations the remainder of the ($500 h^{-1}$ Mpc)$^{3}$ Millennium volume is also simulated, but with dark matter only and at much lower resolution.  This ensures that the influence of the surrounding large-scale structure is accurately accounted for.

\subsection{Sample selection}

We use the same sample of 412 Milky Way-mass disc galaxies introduced in Paper I.  A Milky Way-mass system is a defined as a system with a present-day total (gas+stars+dark matter) mass within $r_{200}$ (i.e., the radius which encloses a mean density of 200 times the critical density of the universe) of $7\times10^{11} M_\odot < M_{200} < 3\times10^{12} M_\odot$.  This range roughly spans the mass estimates in the literature for both the Milky Way and M31, although we are not attempting to reproduce the properties of either of these systems in detail.  Systems and their substructures are identified in the simulations using the \subfind\ algorithm of \citet{dolag09}, that extends the standard implementation of \citet{springel01} by including baryonic particles in the identification of self-bound substructures.  For each FoF system that is identified, a spherical overdensity (SO) mass with $\Delta = 200$ is computed (i.e., $M_{200}$).  The gas+stars+dark matter associated with the most massive subhalo of a FoF system are considered to belong to the galaxy, while the gas+stars+dark matter associated with the other self-gravitating substructures are classified as belonging to satellite galaxies.  As the current study is focused on examining the extended stellar distributions of normal disc galaxies, we exclude all bound substructures (i.e., satellites) from the analyses presented in this paper.  We also exclude mass that is part of the FoF system but which is not bound to it (the so-called FoF ``fuzz''), but note that these particles make a negligible contribution to the stellar spheroid by mass.

Galaxies identified in this way are then classified morphologically as disc- or spheroid-dominated, based on their dynamics.  In particular, we use the ratio of disc stellar mass to total stellar mass (D/T) within 20 kpc computed by C10 to select our sample.  In particular, we select disc galaxies with $D/T > 0.3$.  We note that C10 decomposed the spheroid component from the disc by first computing the angular momentum vector of all the stars within 20 kpc, calculating the mass of stars that are counter-rotating, and doubling it.  This procedure therefore assumes that the spheroid has no net angular momentum.  The remaining mass of stars is assigned to the disc and D/T is computed as the disc mass divided by the sum of the disc and spheroid masses.  In reality, however, the spheroid of the simulated galaxies (and likely of real galaxies as well) does in general possess a net angular momentum.  We explicitly show this below for the simulated galaxies.  However, ignoring this fact in the {\it selection} of our sample has no important consequences for our results or conclusions, since the vast majority of our simulated galaxies are highly disc-dominated regardless of whether or not we take into account the angular momentum of the spheroid when calculating D/T.

\begin{figure*}
\includegraphics[width=170mm]{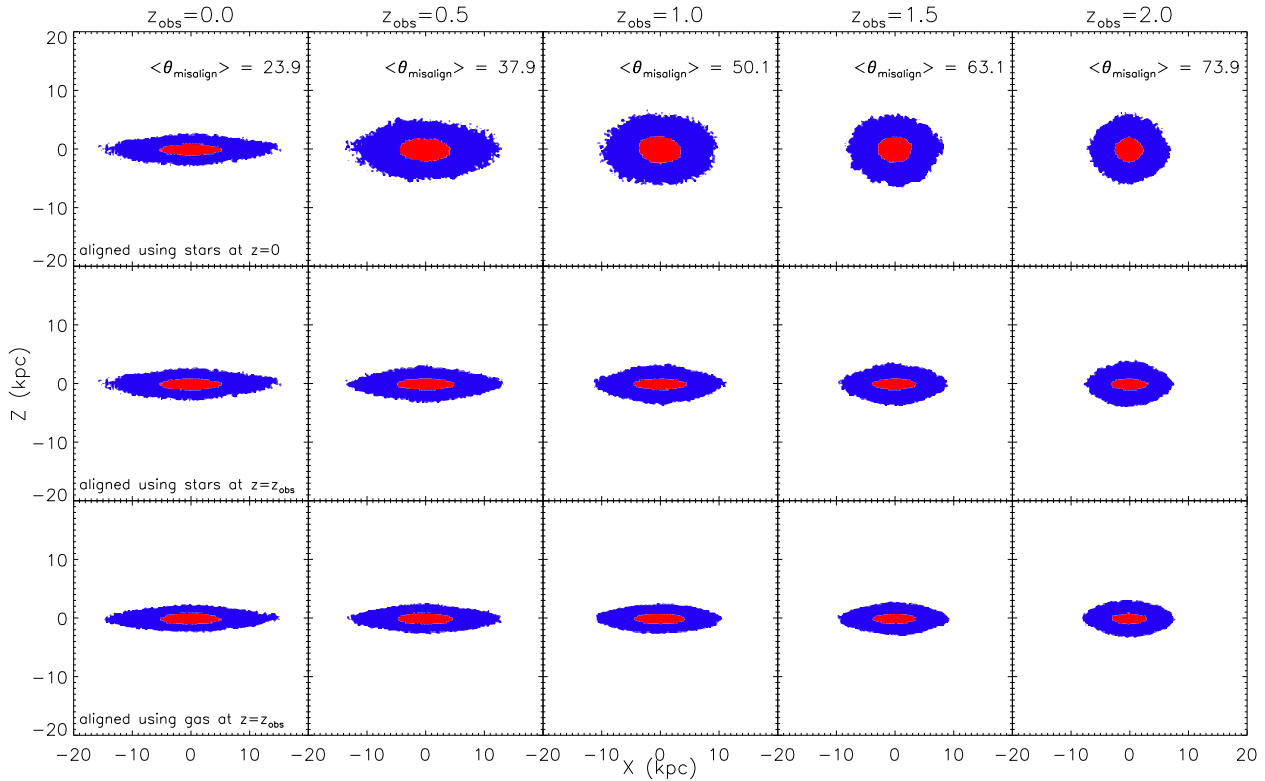}
\caption{\label{fig:disc_stack}
Stacked spatial distributions (using all 412 galaxies) of the cold, star forming gas at various redshifts $z_{\rm obs}$.  Only gas in the most massive progenitor subhalo is shown (i.e., in situ star forming gas).  In the top row the individual galaxies have been aligned before stacking so that the angular momentum vector of their stars at $z=0$ is parallel to the Z-axis.  The median misalignment angle between the cold gas at $z=z_{\rm obs}$ and the stars at $z=0$ is quoted in the legend.  In the middle and bottom rows the individual galaxies are aligned using the $z=z_{\rm obs}$ angular momentum vector of the stars and cold gas, respectively, before stacking.  The red and blue filled contour enclose 50\% and 90\% of the star forming gas particles, respectively.  The vast majority of in situ stars formed in a disc. The median misalignment between the orientation of the disc at high redshift (where the majority of the in situ stars in the present-day spheroid formed, see Fig.~\ref{fig:tform_hist}) and the disc today is large, implying that most discs have been severely torqued between $z \sim 2$ and the present day.
}
\end{figure*}

The method of C10 for decomposing the disc and spheroid does not assign individual star particles to either component, it simply computes the {\it total} mass in two components.  We follow the approach of \citet{abadi03} in assigning particles to the disc and spheroid.  In particular, we align the total angular momentum of the stellar disc with the Z-axis, calculate the angular momentum of each star particle in the X-Y plane, $J_Z$, and compare it with the angular momentum, $J_{\rm circ}$, of a star particle on a co-rotating circular orbit with the same energy.  Disc particles are then selected using a cut of $J_Z/J_{\rm circ} > 0.8$ as in \citet{zolotov09}.  Visually, this cut is quite successful in isolating the disc, but in a number of cases we found that particles well above/below the disc mid-plane were also assigned to the disc.  For this reason, we also apply a spatial cut such that disc particles cannot exist more than $2 h^{-1}$ kpc above or below the disc mid-plane (in Fig.~\ref{fig:disc_stack} below we show that star-forming gas in the disc is generally confined to this region).  We have also tried assigning disc particles based on the fraction of stellar kinetic energy in ordered rotation, as defined in \citet{sales10} (see also Fig.~\ref{fig:krot_track} below), but found no significant differences from the method outlined above.

Finally, we point out that we have not imposed any explicit constraints on the merger histories of the simulated galaxies.  This is in contrast with many of the previous cosmological modelling studies of Milky Way-type galaxies, which usually select their simulation candidates from systems that have not undergone any major mergers in the recent past (e.g., \citealt{cooper10}).  As noted in Paper I, the constraints adopted in previous hybrid studies are overly conservative, given that the majority of the Milky Way-mass systems in \gimic\ are disc-dominated systems (see C10).  The resilience of galactic discs to mergers likely stems from the fact that the galaxies contain a significant dissipational gaseous component, which is able to maintain (or re-establish) the disc (see, e.g., \citealt{hopkins09,moster11}).  In any case, for the present study, which is based on hydrodynamic simulations of large volumes, there is no need to place explicit constraints on the merger histories since it is easy to identify which galaxies have prominent discs at $z=0$.

Applying the system mass and morphology criteria described above yields a total sample of 412 systems in the 5 intermediate-resolution \gimic\ volumes.

\subsection{Distinguishing accretion and in situ formation}

We define in situ star formation as star formation that takes place in the most massive subhalo of the most massive progenitor (MMP) FoF group of the present-day system.  Stars are said to be `accreted' if they formed either in another FoF group or in a satellite galaxy (i.e., not the dominant subhalo) of the MMP.  In practice, only a small proportion (2\%) of the stellar halo is from stars that formed in satellites of the MMP and were then stripped; most of the accreted stars were formed in other FoF groups (i.e., other haloes) before they became satellites of the MMP.  

To identify the MMP of a given Milky Way-mass system at $z=0$, we first select all dark matter particles within $r_{200}$ belonging to the dominant subhalo.  We then use the unique IDs assigned to these particles to identify the FoF group in previous snapshots that contains the largest number of these particles.  This earlier  FoF group is defined to be the MMP and the dominant subhalo of that FoF group is assumed to be the most massive progenitor of the dominant subhalo of the system at $z=0$.
Each of the star particles that comprise the $z=0$ stellar spheroid is tracked back in time to the FoF group and subhalo it belonged to, if any, when it formed from a gas particle.  
If, at the time of star formation, the star particle is a member of the dominant subhalo of the MMP, the star particle is said to have formed in situ.  If it formed in another FoF group or in a non-dominant subhalo of the MMP, then it is said to have been accreted.

As discussed in Paper I, there are sufficient snapshots output for each of the \gimic\ volumes to be able to reliably assign particles to a formation mode. We trace each of the $z=0$ stellar halo particles back in time as far as $z=10$ to determine whether they formed in situ or were accreted. 

\section{Origin of the in situ component}

Before examining the shape and kinematics of the stellar spheroid, we begin by examining in some detail the origin of the in situ component, which dominates the spheroid by mass in our simulations.  We show below that there is a close physical connection between the in situ component of the spheroid and the early disc.  This motivates us to look at potential differences in the shape and kinematics of the in situ and accreted components of the spheroid in Section 4.

In Fig.~\ref{fig:disc_stack} we show the stacked (using all 412 disc galaxies) spatial distribution of cold, dense star forming gas ($T < 10^5$ K and $n_{\rm H} > 0.1$ cm$^{-3}$) at various redshifts, $z_{\rm obs}$.  We only include gas that is part of the most massive subhalo of the most massive progenitor of the disc galaxies that were selected at $z=0$.  This is, by definition, the gas that gives rise to the in situ stellar component of the galaxy (both of the disc and the spheroid).  The red and blue filled contour enclose 50\% and 90\% of the star forming gas particles, respectively.

In the top row we show the stacked distributions where the individual galaxies have been aligned before stacking so that the angular momentum vector of their stars at $z=0$ is parallel to the Z-axis.  At $z=0$ (top left panel) a disky distribution is evident.  This implies that the bulk of the star formation is occurring in a disc and that the star forming gas disc is well-aligned with the stellar disc.  As we track the most massive progenitor of these galaxies back to previous redshifts and stack using the {\it same ($z=0$) rotation matrix} to align the gas the stacked star forming region becomes less disky and more isotropic in appearance.  Either the star forming gas was genuinely distributed in this way (and therefore discs only formed very recently), or the gas was in a disc but the disc has been significantly torqued between $z=0$ and $z=z_{\rm obs}$.  

We can distinguish between these two possibilities by aligning the star forming gas at $z_{\rm obs}$ using either the angular momentum of the stars (middle row) or the angular momentum of the cold gas itself (bottom row) at $z=z_{\rm obs}$.  A disky distribution for the star forming gas is readily apparent using either approach.  The radial extent of the disc becomes smaller at high redshift which is qualitatively consistent with previous simulation studies of disc galaxies as well as with observations (e.g., \citealt{wang11}).  Thus, the star forming gas becomes more isotropic looking with increasing redshift in the top row because of the significant torquing of the disc.  In the legend in the top row we quote the median misalignment angle (in deg.), $\theta_{\rm misalign}$, between the angular momentum vector of the star forming gas at $z=z_{\rm obs}$ and that of the stars at $z=0$.  A random distribution (i.e., no correlation between the vectors) would have a median misalignment of 90 degrees.  This is nearly the case of the star forming discs at $z=2.0$, implying that there is almost no correlation between the orientation of (proto)-discs at $z=2.0$ and the descendant discs today. 

Based on Fig.~\ref{fig:disc_stack} we therefore conclude that the vast majority of the overall (disc+spheroid) in situ component present in our galaxies at $z=0$ formed in a disc at some previous time.  Since stars formed in situ dominate the stellar spheroid in these simulations out to $r_{\rm 3D} \sim 20-30$ kpc (see Fig.\ 7 of Paper I), this implies that some process has added sufficient energy (`puffed up') to the stars to liberate them from the confines of the disc.  We can explicitly check that this is the case by comparing the ratio of (specific) kinetic energy in ordered rotation [$K_{\rm rot} \equiv (J_Z/R)^2/2$, where $R = (X^2 + Y^2)^{1/2}$] to total kinetic energy [$K_{\rm tot} \equiv v_{\rm tot}^2/2$, where $v_{\rm tot} \equiv (v_X^2+v_Y^2+v_Z^2)^{1/2}$] for in situ stars at the present-day to the same ratio for the same particles at the time of star formation.  Operationally, we select cold, star forming (in situ) gas particles residing in the most massive progenitor at some previous redshift $z_{\rm obs}$ and then find the corresponding star particles at $z=0$ using the unique particle IDs stored in the snapshot data.  Note that not all of the cold, star forming gas turns into stars by $z=0$ (e.g., some gas is heated up by SNe) and not all of the in situ stars at $z=0$ will be in the cold gas component at $z_{\rm obs}$.  We are therefore examining a subset of the particles, but it is an unbiased selection in terms of kinematics.

\begin{figure}
\includegraphics[width=\columnwidth]{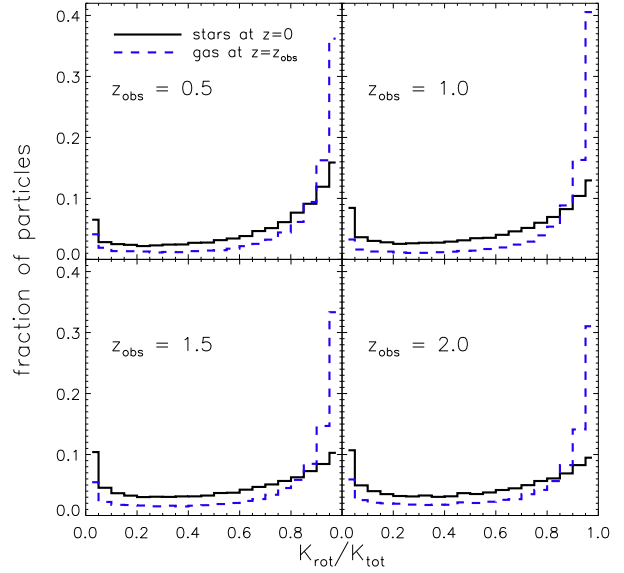}
\caption{\label{fig:krot_track}
The fraction of kinetic energy in ordered rotation for in situ stars at $z=0$ and the cold gas which formed them at $z=z_{\rm obs}$.  The different panels show different values of $z_{\rm obs}$.
The specific kinetic energy in ordered rotation is $K_{\rm rot} \equiv (J_Z/R)^2/2$ (where $R = (X^2 + Y^2)^{1/2}$) and the total specific kinetic energy is $K_{\rm tot} \equiv v_{\rm tot}^2/2$ (where $v_{\rm tot} \equiv (v_X^2+v_Y^2+v_Z^2)^{1/2}$).  Operationally, cold, star forming (in situ) gas particles are selected at $z_{\rm obs}$ and the corresponding star particles at $z=0$ are found using the unique particle IDs.  At any redshift $z_{\rm obs}$ the cold, star forming gas has a much larger fraction of its energy in highly ordered rotation (e.g., $K_{\rm rot}/K_{\rm tot} > 0.8$) than the stars at $z=0$ which formed from that gas.  This difference increases with increasing redshift.
}
\end{figure}

The results of this test are shown in Fig.~\ref{fig:krot_track}.  At any redshift $z_{\rm obs}$ the cold, star forming gas has a much larger fraction of its energy in highly ordered rotation (e.g., $K_{\rm rot}/K_{\rm tot} > 0.8$) than the stars at $z=0$ which formed from that gas.  As we look at cold gas further back in time (increasing $z_{\rm obs}$), we see that the stars at $z=0$ that formed from that gas have lower and lower fractions of their energy in highly ordered rotation.  This is what one would expect, as there has been more time for the stars to be dynamically puffed up (e.g., via accretion, minor mergers, disc instabilities, and so on).  

\begin{figure}
\includegraphics[width=\columnwidth]{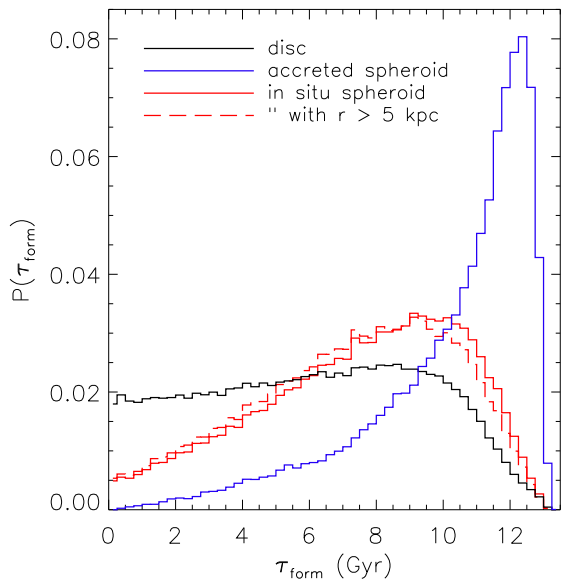}
\caption{\label{fig:tform_hist} 
Distribution of stellar ages.  Most of the accreted component of the spheroid is very old (typically 11-12 Gyr).  The disc is younger and has roughly had a constant SFR for the past $\approx 9-10$ Gyr.  The in situ spheroid component has a slightly older median age than the disc.  The SFR of this component has been declining steadily since $z \approx 1.5$.  Note that the amplitudes of the histograms are determined by normalising to the total number of stars in {\it each} population separately.
}
\end{figure}

When did the stars in the in situ spheroid form?  In Fig.~\ref{fig:tform_hist} we show the stacked distribution of stellar ages of the disc and the in situ and accreted spheroid components at $z=0$.  As expected, the accreted spheroid is old (median age of $\approx 11.1$ Gyr, or $z \approx 2.7$), having been built up from the tidal disruption of dwarf galaxies.  As shown in Paper I, however, the accreted spheroid is still being {\it assembled} today (see Fig.\ 10 of that study).  Both the in situ spheroid and disc are on average younger than the accreted spheroid.  Interestingly, the median age of the in situ spheroid ($\approx 8.0$ Gyr, corresponding to $z \approx 1.1$) is significantly older than the median age of stars in the present-day disc ($\approx 6.2$ Gyr, or $z \approx 0.7$).  Taken together with Figs.\ 1-2, this implies that the in situ component of the spheroid primarily formed in a {\it proto-}disc that evidently was heated up before much of the current disc stellar population was formed. 

Interestingly, there is a marked decline in the distribution of the in situ spheroid toward younger ages, even though the distribution of ages for the disc is effectively flat below $\approx$ 9-10 Gyr (implying a nearly constant disc star formation rate for the past $\approx$ 9-10 Gyr).  This suggests that the heating mechanism that is responsible for producing the in situ spheroid is becoming less and less efficient at late times.  A likely culprit for the heating is the accretion of satellites (and mass in general) onto the simulated galaxies.  Recently, \citet{boylan-kolchin10} studied the mass accretion histories of thousands of Milky Way-mass dark matter haloes in the Millennium-II simulation.  If one defines the epoch of formation as the time when the halo has accreted half of its present-day mass (e.g., \citealt{lacey93}), \citet{boylan-kolchin10} find a median epoch of formation of $z \approx 1.2$ for Milky Way-mass haloes.  This is suggestively close to the median age of the in situ spheroid component.  Alternatively, instead of using the epoch of formation, one can use the redshift of most massive merger.  \citet{boylan-kolchin10} find a median mass ratio of 10:1 for the most massive merger (i.e., the mass of the satellite at accretion is 10\% of the final virial mass of the main halo).  We have tracked in the \gimic\ simulations all of the galaxies in our sample which have undergone a merger with mass ratio 10:1 (or lower).  The median redshift for systems with such mergers is $\approx 1.1$.  Both arguments point to dynamically active systems (on average) at this period of time\footnote{The lack of a correlation between the orientation of the present-day disc and proto-disc in which the in situ spheroid stars were formed (evident in Fig.~\ref{fig:disc_stack}) adds further evidence of a dynamically active state at $z \sim 1$.}.  After $z \sim 1$ the mass accretion histories flatten significantly (see Fig.\ 2 of \citealt{boylan-kolchin10}), which is qualitatively consistent with the decline in the stellar age distribution of the in situ spheroid.

\begin{figure}

\includegraphics[width=\columnwidth]{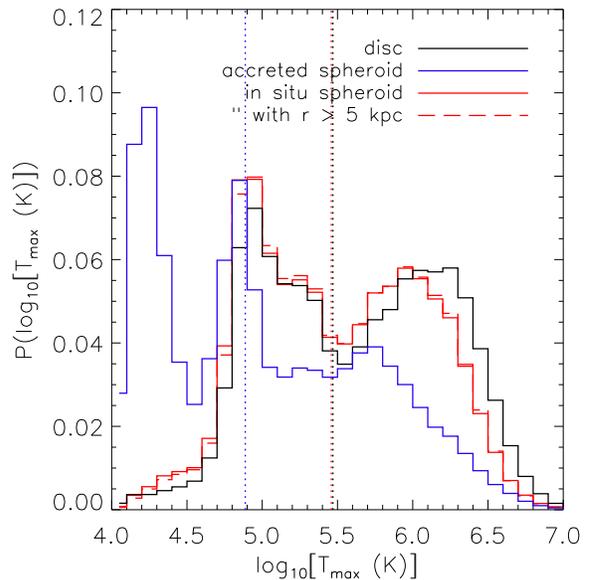}
\caption{\label{fig:Tmax_hist} 
Distributions of maximum past temperatures of the gas which formed the disc and the accreted and in situ spheroid components.  The black, blue, and red dotted vertical lines correspond to the median $T_{\rm max}$ for the disc and the accreted and in situ spheroid components, respectively.
The accreted spheroid formed mostly from cold gas (``cold mode''), whereas approximately half of the stars in the disc and half in the in situ spheroid formed from gas that was accreted in the so-called ``hot mode'' (i.e., they formed from the cooling of gas that was first shock heated to approximately the virial temperature).  
}
\end{figure}

Our analysis therefore favours a scenario similar to that proposed recently by \citet{purcell10}, in which the inner parts of galactic spheroids contain proto-disc stars that may have been liberated in the same merger/accretion events that delivered material to the outer spheroid.  Note that \citet{purcell10} primarily analysed idealised collisionless simulations of disc bombardment by infalling satellites, whereas \gimic\ is a suite of fully cosmological hydrodynamic simulations.

Finally, we make a brief comment on the thermodynamic state of the gas from which the in situ component formed.  In Fig.~\ref{fig:Tmax_hist} we show the distribution of the maximum past temperature of the gas particles from which the disc, in situ spheroid and accreted spheroid star particles formed.  The distributions are obviously bimodal.  The distributions for the disc and in situ spheroid components are quite similar, with a peak\footnote{This lower $T_{\rm max}$ peak corresponds to the temperature floor imposed by the photoionising UV/X-ray background in the simulations.  Note that the temperature floor that is established depends on redshift, as the cooling rate of the gas increases with redshift due to the increase in density with redshift.  This is why the lower $T_{\rm max}$ peak for the accreted component, which formed at higher redshift, is only $\approx 10^{4.3}$ K (see \citealt{vandevoort11}).} at $T_{\rm max} \approx 10^5$ K and another at $T_{\rm max} \approx 10^{6.0-6.3}$ K, corresponding to so-called ``cold mode'' and ``hot mode'' accretion (e.g., \citealt{birnboim03,keres05,keres09,vandevoort11}), respectively (note that the virial temperature for haloes we are studying here is $T_{\rm vir} \sim 10^6$ K).  In agreement with C10, we find that approximately half of the disc stellar mass formed via the cooling of hot gas.  We find a similar result for the in situ spheroid component.  By contrast, the accreted component formed predominantly from gas that has never been hot.

The above points to a different physical nature for the in situ component than advocated previously by \citet{zolotov09}.  In particular, these authors found a median redshift of formation of $\approx 3$ for the in situ spheroid component in their simulations.  Furthermore, the stars formed at very small radii ($\la 1$ kpc) from gas that was fed via cold mode accretion.  In contrast, we find the in situ component is younger, formed in a proto-disc, and that approximately half of the stellar mass was formed via the cooling of gas that was heated to approximately the virial temperature of the system.  A possible reason behind this difference is that the simulations adopted by \citet{zolotov09} suffered from a larger degree of overcooling due to inefficient feedback.  A likely consequence of this overcooling is that a significant fraction of the gas that would have been destined to end up as part of the in situ stellar halo was instead converted into stars in small dwarf galaxies at high redshift, and were therefore instead incorporated into the accreted spheroid component when these systems fell into the main progenitor.  This scenario is consistent with the fact that the simulations used by \citet{zolotov09} produce present-day satellite galaxies that are too bright compared to those of the Milky Way or M31 (see Fig.\ 4 of \citealt{governato07}).  We note, however, that there are many other differences between the simulations that could also play a role (e.g., somewhat different implementations of star formation, chemodynamics, and metal-dependent radiative cooling).  Direct comparisons of the properties of stellar spheroids predicted by different simulation codes all using the same initial conditions would obviously be beneficial in helping to firmly identify the origin of the apparent differences between the simulations.

The above results point to a close physical connection between the disc and the in situ spheroid in our simulations. Naively, we therefore expect that the latter should have kinematics and a spatial configuration that may be more reminiscent of the former than to the accreted spheroid.  As we highlighted in Section 1, there is mounting observational evidence for net rotation and flattening of the inner spheroid of both the Milky Way and M31.  This is a tantalising suggestion that the inner halo indeed contains a significant dissipational component.  Below we study the shape and kinematics of the stellar `spheroid' and, where possible, make comparisons to the observational data.

\section{Structure and Rotation}

\subsection{Structure}

\begin{figure*}
\includegraphics[width=\columnwidth]{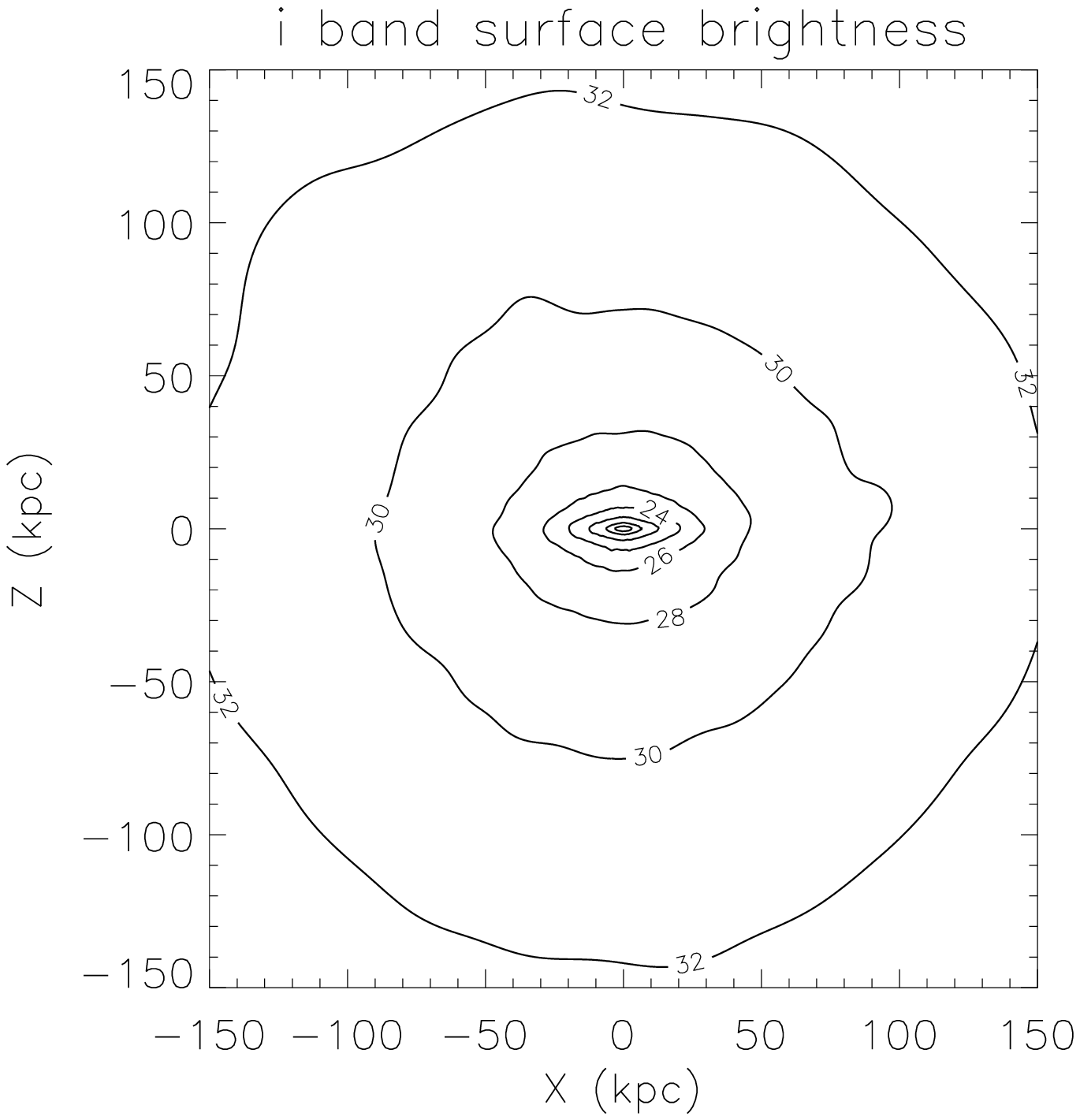}
\includegraphics[width=\columnwidth]{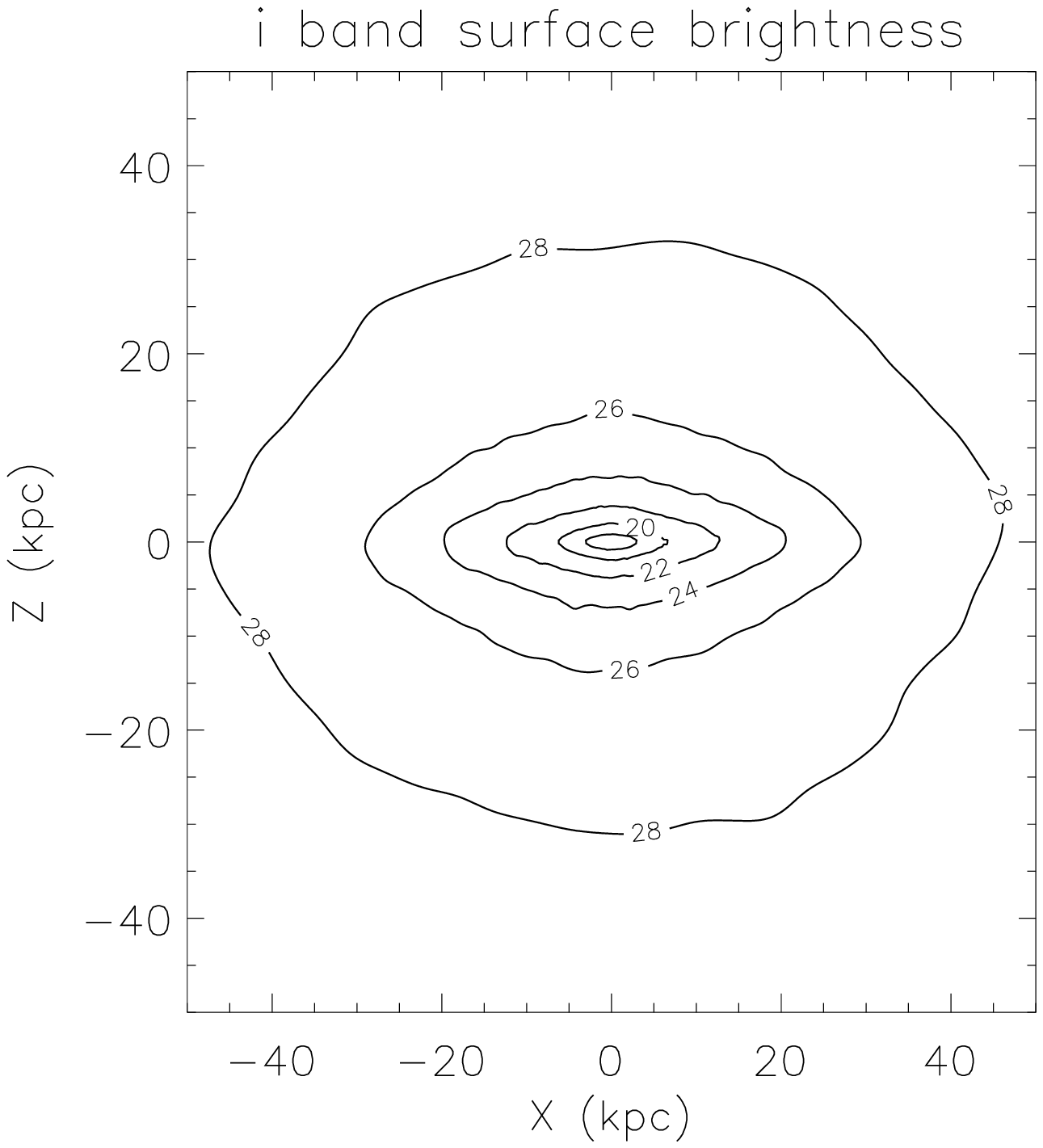}
\caption{\label{fig:shape_iband}
Stacked SDSS i-band surface brightness contour maps of the simulated galaxies.  The galaxies are rotated to an edge-on configuration before stacking.  Self-gravitating substructures (satellites) have been excised and the map has been heavily smoothed using an SPH kernel (see text for details).  The numbers associated with the contours give the mean surface brightness in mag arcsec$^{-2}$.  {\it Left:} Large-scale view.  {\it Right:} Zoom in on the central regions.  Note the stellar disc is generally confined to $|Z| < 2$ kpc.  It is readily apparent that the spheroid shows a significant degree of flattening, particularly in the inner regions ($r \la 40$ kpc).
}
\end{figure*}

Below we examine the structure of the spheroid from external and internal perspectives separately.  We make comparisons to measurements of M31 and NGC 891 and stacked observations of more distant galaxies when considering the former and to the Milky Way when considering the latter.

\subsubsection{Structure - External view}

We begin by examining the average 2D stellar distribution of the spheroid.  To this end, we stack all 412 simulated galaxies in an edge-on configuration, by first aligning the angular momentum vector of the stars of each galaxy with the Z-axis of the simulation box.  Since we are interested in the diffuse spheroid, we stack only the star particles that are bound to the most massive subhalo (the main galaxy; i.e., we exclude star particles bound to orbiting satellite galaxies).  To facilitate easy comparisons with observations (see below), we compute a stacked SDSS i-band surface brightness map\footnote{The i-band luminosities are computed by treating each star particle as a simple stellar population (SSP).  The simulations adopt a Chabrier IMF and store the age and metallicity of the particles.  We use this information to compute a spectral energy distribution for each star particle using the GALAXEV model of \citet{bruzual03}.  The i-band luminosity is obtained by integrating the product of the SED with the SDSS i-band transmission filter function.  Our calculation neglects the effects of dust attenuation, which may be relevant at very small radii.}.  The particles are interpolated to a 2D regular grid after stacking using the smooth particle hydrodynamics (SPH) kernel with a large number of smoothing neighbours (512).  This was done to smooth out substructures not associated with self-bound satellites (again we are interested in the global/average properties).  We have checked that using smaller numbers of smoothing neighbours does not affect the average shape as a function of projected radius, it does however result in a more noisy profile.

The resulting smoothed i-band map is shown in the left panel of Fig.~\ref{fig:shape_iband}.  The numbers associated with the contours give the mean surface brightness in mag arcsec$^{-2}$.  A zoom in on the central 50 kpc is shown in the right panel.  Note that the disc stars are located at $|Z| < 2$ kpc (see Fig.~\ref{fig:disc_stack}).  It is readily apparent that the spheroid component shows a significant degree of flattening and that the flattening is much more pronounced in the inner regions.

\begin{figure}
\includegraphics[width=\columnwidth]{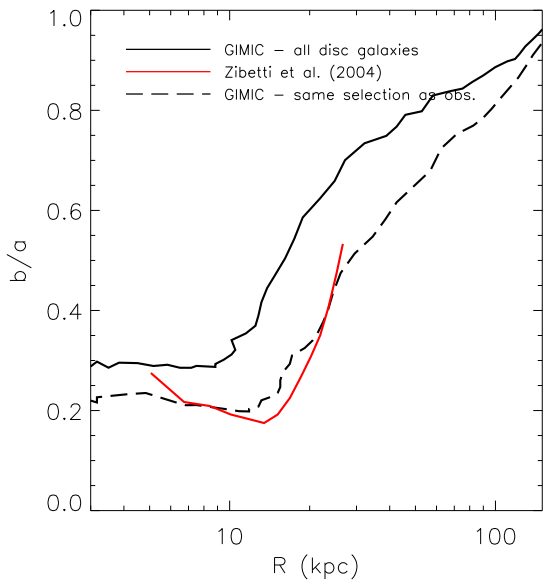}
\caption{\label{fig:zibetti}
The flattening ($b/a$) of the average surface brightness distribution as a function of distance along the major axis ($R$). The solid red curve represents the flattening profile derived by \citet{zibetti04} from a stacking analysis of $\approx 1000$ edge-on disc galaxies in the SDSS.  The solid black curve is the flattening profile derived from an identical stacking analysis of all 412 simulated disc galaxies.  The dashed black curve represents the average flattening profile when we stack only galaxies that would make the sample selection criteria of \citet{zibetti04} (see text).  The flattening profile of the simulated galaxies bears a remarkable resemblance to that derived from observations.
}
\end{figure}

\citet{zibetti04} have performed a stacking analysis of approximately 1000 edge-on disc galaxies in the SDSS and have measured the surface brightness distribution of the spheroid down to $\sim 30$ mag arcsec$^{-2}$ in the i-band.  We can directly compare our results above to this study.  \citet{zibetti04} have quantified the degree of flattening by fitting ellipses to their stacked data at a variety of surface brightness levels.  Their trend of $b/a$ (where $b$ and $a$ are the semi-minor and semi-major axes respectively) with distance along the major axis\footnote{We have converted their distances from pixel lengths into kpc by adopting the mean redshift of their sample, which is $z=0.05$.} is shown as the solid red curve in Fig.~\ref{fig:zibetti}.   We have performed an identical analysis to the stacked i-band image of all 412 simulated galaxies, which is represented by the solid black curve.  The dashed black curve represents the average flattening profile when we stack only galaxies that would make the sample selection criteria of \citet{zibetti04}, namely that the isophotal semi-major axis $a$ exceed 10 arcseconds and that the isophotal $b/a$ be small in order to select disc galaxies\footnote{Isophotal values of $a$ and $b$ refer to the values produced by the SDSS reduction pipeline at a surface brightness level of 25.0 mag arcsec$^{-2}$.}. \citet{zibetti04} adopt an isophotal $b/a$ threshold of $< 0.25$, whereas we adopt a slightly larger threshold of $0.4$ in order to boost our sample of galaxies.  Only 41 galaxies in our sample have $b/a < 0.25$, whereas approximately half (210) have $b/a < 0.4$.  Imposing a cut of $b/a < 0.25$ gives results that are consistent with (but considerably more noisy than) our default cut of $b/a < 0.4$.

Overall there is good agreement in the shape of the flattening profile between the simulations and the measurements of \citet{zibetti04}.  This statement is independent of whether or not we apply the same selection criteria for the disc galaxy sample as \citet{zibetti04}.  When we apply the same selection criteria, however, this results in an improved match to the normalisation.  At $r \la 10$ kpc the profile asymptotes to a level $b/a \approx 0.2$.  This is due to the presence of the stellar disc.  Beyond this radius we begin to probe into the spheroid.  Even out to radii of $50$ kpc the spheroid is still significantly flattened ($b/a \approx 0.7$).  

The agreement between the simulations and observations is non-trivial, as it has proved notoriously difficult to produce reasonable disc galaxies at all in cosmological hydrodynamic simulations.  As discussed in Paper I, part of the reason for the improvement with \gimic\ is that it does not suffer from strong overcooling for Milky Way-mass (and lower) galaxies owing to efficient (but energetically feasible) SN feedback.  The feedback parameters themselves have been tuned only to match the peak of the global star formation rate density of the universe.  As noted in the Introduction, strong SN feedback is also required to explain the low baryon fractions of normal galaxies and the enrichment of the intergalactic medium.

In the above we have focused on the average shape of the spheroid that would be derived by stacking large samples of galaxies.  This was required to probe the shape of the outer spheroid (beyond $r \ga 30-40$ kpc), since our simulations contain too few particles to probe the shape to very large radii on a system-by-system basis.  As we will show later, however, the simulations predict that there should be a significant degree of system-to-system scatter in both the shape and kinematics of the inner spheroid, where observations are generally currently limited to anyway.  At present it is quite challenging to probe the spheroids of individual galaxies, owing to the very low surface brightness of this component.  While in the future this will be readily achievable for large numbers of galaxies with, e.g., the {\it LSST}, studies of individual galaxies are presently limited to the local volume.  In the case of M31, \citet{pritchet94} find that the spheroid is significantly flattened with $b/a \approx 0.55$ at a minor axis distance of 10 kpc.
More recently, \citet{ibata05} have found evidence for an extended disc-like structure out to $\sim 50$ kpc which is aligned with the stellar disc and has $b/a \approx 0.6$. 
  The large panoramic view of M31 to be provided by the ongoing Pan-Andromeda Archaeological Survey (PAndAS) will be useful for quantifying the degree of flattening as a function of distance out to very large radii.  The flattening of the spheroid of NGC 891, a local Milky Way analogue, has been measured by \citet{ibata09}.  They find a flattening of $b/a \approx 0.5$ out to a distance of $\sim 25$ kpc.  Both M31 and NGC 891 therefore show a level of flattening that is comparable to the mean value derived from our simulated galaxies.

\begin{figure}
\includegraphics[width=\columnwidth]{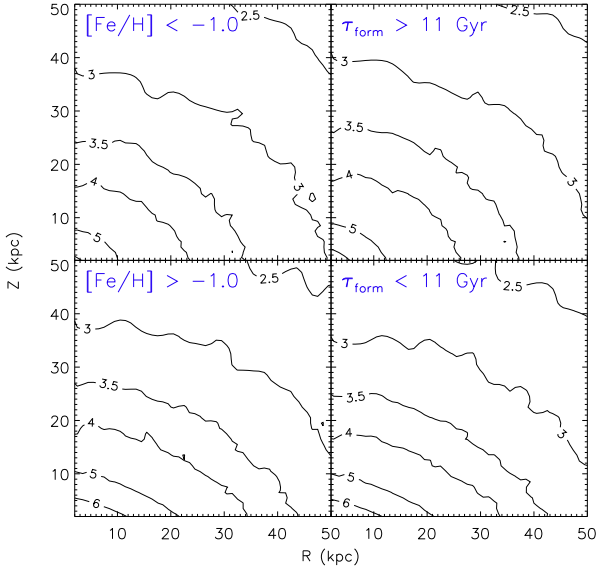}
\caption{\label{fig:shape_r_z}
Stacked stellar mass density contours in cylindrical coordinates. Self-gravitating substructures (satellites) have been excised and the map has been heavily smoothed using an SPH kernel (see text for details).  The contours reflect the logarithm of the stellar mass density in $M_\odot$ kpc$^{-3}$ (for the spherically-averaged stellar mass density refer, to Fig.\ 3 of Paper I).  The various panels show the density distribution for different cuts in the metallicity and age of the star particles.  A comparison of the top and bottom panels shows that the flattening of the stellar mass density is much more pronounced for younger and more metal-rich stars and at smaller radii, in qualitative agreement with the measurements of the Milky Way by \citet{carollo07}.
}
\end{figure}

\subsubsection{Structure - Internal view}

Measurements of the spatial structure (and substructure) and kinematics of the spheroid of our own Milky Way galaxy have rapidly increased in quantity and precision in recent years, particularly with the advent of the SDSS, and are set to take another leap in the coming years with the release of data from SDSS-III/SEGUE-2 (Rockosi et al.\ in prep), RAVE, SkyMapper, Pan-Starrs and the launch of Gaia.  Some of the best constraints on the formation and evolution of spheroids and galaxies in general will therefore come (and are now coming) from our own Galaxy.  With this as motivation, we have examined the structure of the spheroid from an internal perspective.

In Fig.~\ref{fig:shape_r_z} we plot stacked stellar mass density contours in cylindrical coordinates.  The maps again exclude self-gravitating substructures (satellites) and have been smoothed to a 2D regular grid (in R,Z space) after stacking using the SPH smoothing kernel with 512 smoothing neighbours.  The contours reflect the logarithm of the stellar mass density in $M_\odot$ kpc$^{-3}$ (for the spherically-averaged stellar mass density, refer to Fig.\ 3 of Paper I).  The various panels show the density distribution for different cuts in the metallicity and age of the star particles [i.e., analogous to maps made of the Milky Way by \citet{carollo07} using SDSS data, see their Fig.\ 5].

A comparison of the top and bottom panels reveals that the flattening of the stellar mass density is much more pronounced for younger and more metal-rich stars and at smaller radii.  Qualitatively this is in good agreement with what \citet{carollo07} find for the Milky Way\footnote{Note we did not attempt to adopt the exact same cuts in metallicity employed by \citet{carollo07} since the simulations adopt empirical yields and Type Ia SNe rates, both of which are uncertain at the factor of 2 level \citep{wiersma09b}.  Thus, the metallicities of star particles in the simulations can be shifted up or down by at least 0.3 dex.}.  By selecting young, metal-rich stars, we are predominantly selecting the stars which formed in situ in the simulated galaxies (old, metal-poor stars, by contrast, were predominantly formed in dwarf galaxies which have subsequently been accreted by the main galaxy and tidally destroyed).  The flattening of the {\it overall} distribution, as clearly seen in Fig.~\ref{fig:shape_iband}, is therefore driven primarily by the in situ component of the spheroid.  The gradual transition from a highly flattened to more spherical configuration with distance from the galaxy centre (as in Fig.~\ref{fig:zibetti}) mainly reflects the increasing importance of the accreted component of the spheroid with increasing radius (see Fig.\ 7 of Paper I).

\begin{figure}
\includegraphics[width=\columnwidth]{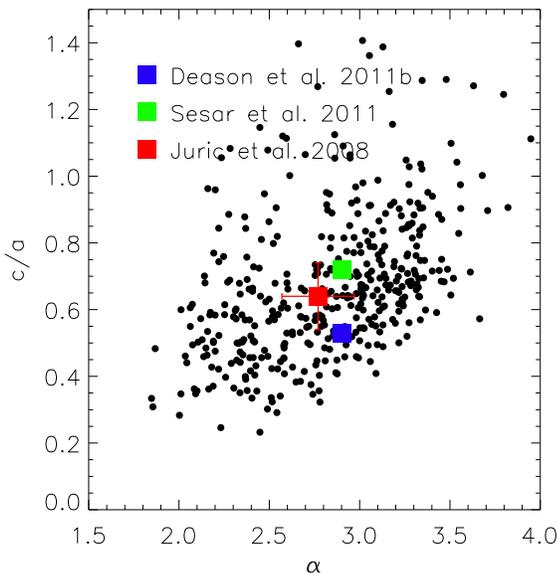}
\caption{\label{fig:shape_only}
Scatter plot of the best-fit parameters (flattening $c/a$ and power-law index $\alpha$) of the oblate power-law (eqn. 1) derived from fitting to each of the simulated galaxies.  For comparison we show recent observational estimates for the Milky Way.  There is a large spread in the parameters but with a clear correlation between the two.  The estimates for the Milky Way are fully compatible with those of the simulated galaxy population, unlike hybrid models which lack any dissipational component and find highly prolate (as opposed to oblate) distributions for the spheroid.
}
\end{figure}

In the above we have considered only the {\it mean} structure of the spheroid, derived by stacking all of our simulated galaxies.  How much system-to-system scatter is there in this structure?  We can assess this by fitting a smooth parametric model to the stellar mass distribution of each of the galaxies and then examining the scatter in the best-fit parameters of that model.  Here we adopt an oblate single powerlaw distribution of the form

\begin{equation}
\rho(R,Z) = \frac{\rho_0}{[R^2 + (Z/q)^2]^{\alpha/2}}
\end{equation}

\noindent where $q \equiv c/a$ and $c$ and $a$ are parallel and perpendicular to the angular momentum vector (Z-axis), respectively.  This distribution has been commonly adopted for modelling the stellar halo of the Milky Way (e.g., \citealt{juric08}).  Note that it has recently been shown that this model does not provide an acceptable fit to the Milky Way's outer halo ($r \ga 25$ kpc, e.g., \citealt{sesar11}; \citealt{deason11b}), in that the observed halo appears to drop off more quickly beyond this radius.  Our simulations lack the mass resolution to probe the detailed shape of the outer halo on a system-by-system basis.  However, as we showed in Paper I, the {\it stacked} mass distribution of our simulated galaxies does indeed drop off faster beyond $r \approx 25$ kpc out to $r \approx 40$ kpc, in rough accordance with observations.  This `kink' in the simulations represents the transition from the in situ-dominated spheroid to the accretion-dominated spheroid.  Beyond $r \approx 40$ kpc the mass density profile asymptotes to a logarithmic slope of $\approx -3.5$.  In any case, for the purposes of the present study, we fit oblate single power-laws to the mass distribution to assess the system-to-system scatter in the simulations.  When comparing to observations, we compare to the best-fit parameters derived by fitting the {\it same} parametric model to the observational data over the same (or similar) range of radii.

When fitting equation (1) to the simulated galaxies, we exclude star particles within $|Z| < 4$ kpc (to exclude disc particles) and beyond a 3D radius $r_{\rm 3D} > 40$ kpc.  This selection\footnote{We find that the best-fit parameters are generally insensitive to the outer radius cut but are somewhat sensitive to the cut in $Z$ if no other cut is applied to exclude disc stars.  In analysis of observational data, cuts on metallicity and/or kinematics are usually adopted to exclude disc stars.  Furthermore, observations, e.g. with the SDSS, are limited in any case to $\theta > 30$ deg, where $\theta$ is the angle between the plane of disc and the height above the plane.} is meant to crudely mimic the typical selection criteria for stars in the halo of the Milky Way.  Finally, before fitting the parametric model, each galaxy is rotated so that the direction perpendicular to the disc plane lies along the Z-axis.  Thus, $c/a < 1$ implies flattening in the same sense as the disc, whereas $c/a > 1$ (which is physically and mathematically possible) would imply the halo is elongated in the direction perpendicular to the disc plane.

\begin{figure*}
\includegraphics[width=\columnwidth]{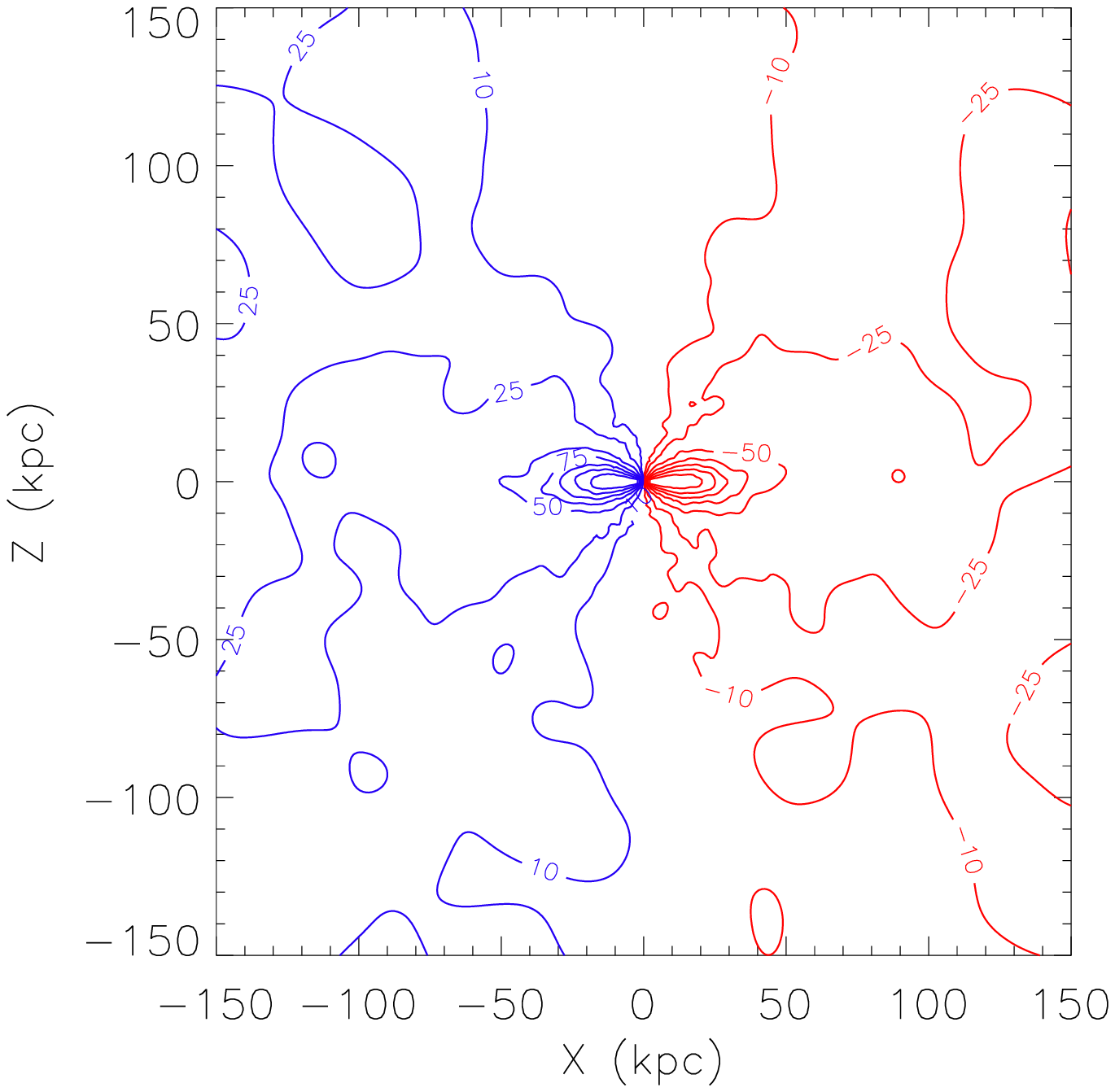}
\includegraphics[width=\columnwidth]{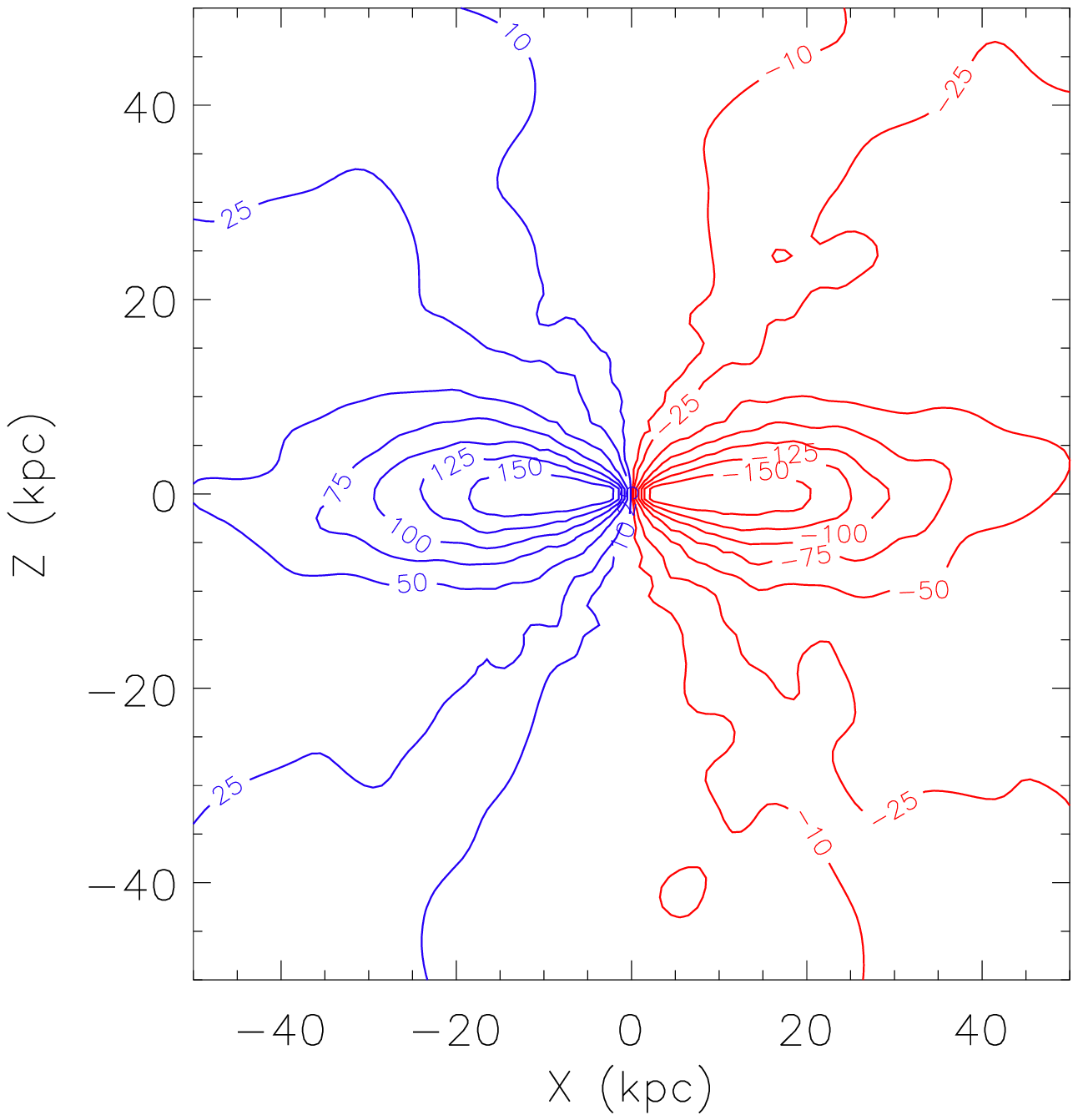}
\caption{\label{fig:rot_stack} 
A stacked line of sight ($v_y$) velocity contour map. All 412 simulated galaxies have been rotated to an edge-on configuration before stacking. Red contours indicate motion away from the observer, whereas blue contours indicate motion towards the observer.  The numbers associated with the contours give the line of sight velocity in km/s.  {\it Left:} Large-scale view.  {\it Right:} Zoom in on the central regions.  Note that the stellar disc is generally confined to $|Z| < 2$ kpc.  The `spheroid' shows a significant degree of rotation, particularly in the inner regions ($r \la 40$ kpc) and bears a close qualitative resemblance to the extended stellar disc around M31 (e.g., \citealt{ibata05}).
}
\end{figure*}

In Fig.~\ref{fig:shape_only} we show a scatter plot of the best-fit parameters $c/a$ and $\alpha$ where each black circle represents a galaxy.  (In Section 4.3 we will explore how these parameters correlate with the rotation of the spheroid and its in situ mass fraction; see Fig.~\ref{fig:shape_rot}.)
It is readily apparent that there is a large range in the values of both the flattening $c/a$ and the power-law index $\alpha$ and that the two are strongly correlated, in that the density drops off more slowly with radius in highly flattened systems.  The flattening $c/a$ ranges from roughly 0.25 to 1.4 with a median of 0.61.  The power-law index ranges from roughly $2$ to $4$ with a median of $2.85$.  (For reference, in Paper I we found that the {\it spherically-averaged} logarithmic slope ranged from roughly $2.5$ to $3.4$ over a similar range of radii.)  For comparison with the Milky Way, we show the best-fit parameters derived by fitting equation (1) to (i) main-sequence turn-off stars in the SDSS \citep{juric08}; (ii) blue horizontal-branch (BHB) stars in the SDSS \citep{deason11b}; and (iii) main-sequence turn-off stars in the CFHTLS \citep{sesar11}.  In spite of the different samples and analysis techniques, similar results are obtained for the spheroid of the Milky Way in all three studies and all are quite compatible with the distribution of spheroid shapes in the simulations.  

\subsubsection{Summary of shape}

Our analysis of the shape of the spheroidal component of the simulated galaxies in \gimic\ indicates that spheroids are oblate in shape (in the same sense as the disc) and that the axial ratios derived from the simulations are consistent with measurements of individual nearby galaxies as well as with stacked observations of more distant (edge-on) galaxies. This is in stark contrast to the results of hybrid models for the formation of stellar haloes, which lack a dissipational component (i.e., the haloes are built from tidal disruption of satellites alone) and typically find a highly {\it prolate} shape for the stellar halo (e.g., \citealt{cooper10}).  The success of the \gimic\ simulations in reproducing the shapes is a direct result of both the large fraction of the spheroid that is produced in situ and quite likely the timing of its formation as well.  We speculate that if the in situ component had formed at very high redshift, well before the formation of a proto-disc, the shape of the spheroid would not be oblate with the observed axis ratio.  

\begin{figure*}
\includegraphics[width=15cm]{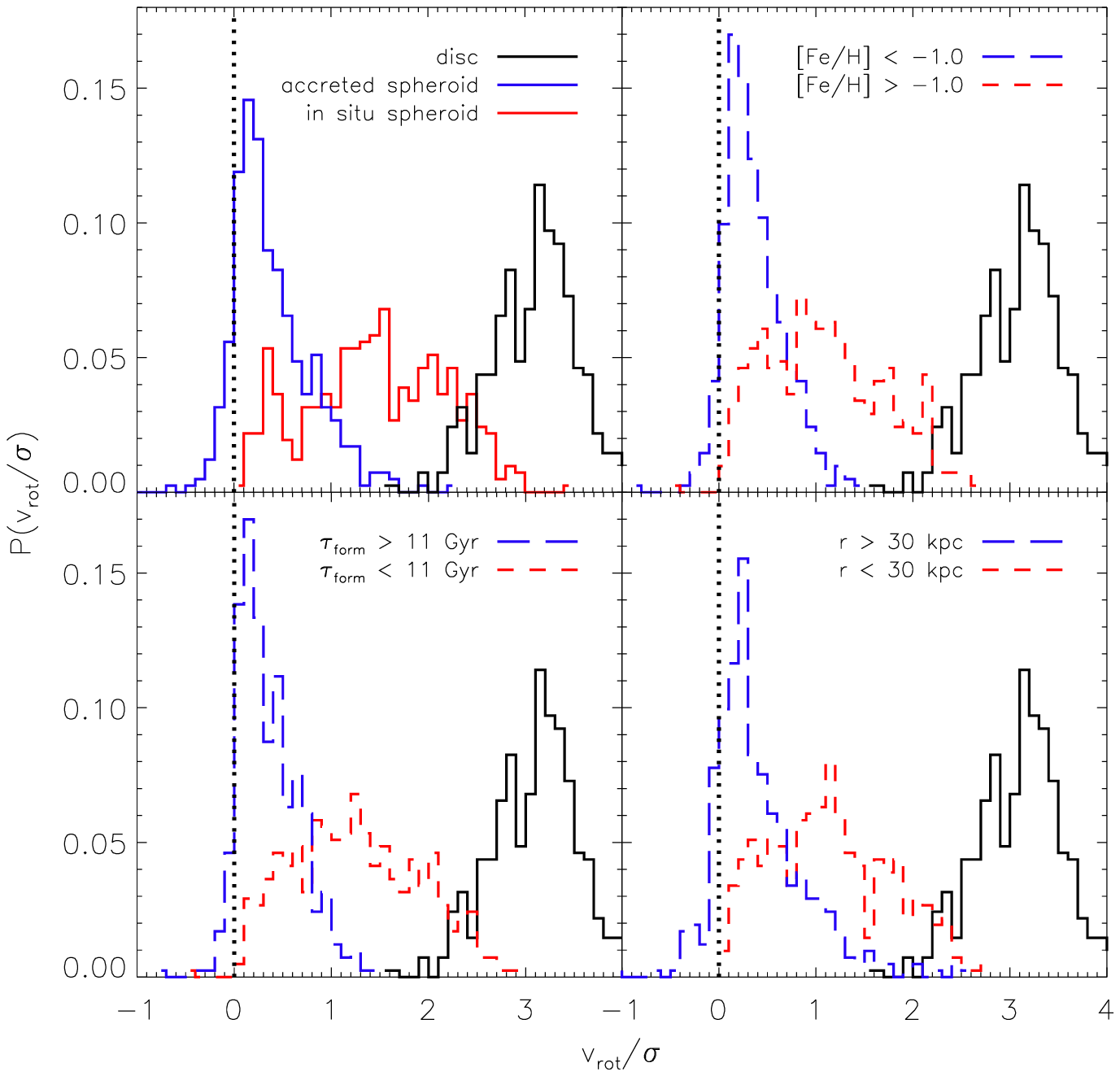}
\caption{\label{fig:vrot_sigma} 
The distribution of the ratio of the rotation velocity to the 3D velocity dispersion of the spheroid of the simulated galaxies (i.e., one value of $V_{\rm rot}/\sigma$ is computed for each simulated galaxy) for different cuts applied to the spheroid.  {\it Top left:} For the spheroid decomposed into in situ and accreted subcomponents.  {\it Top right:} For the spheroid decomposed into relatively metal-rich and metal-poor subcomponents.  {\it Bottom left:} For the spheroid decomposed into relatively young and old subcomponents.   {\it Bottom right:} For the spheroid decomposed into inner and outer subcomponents.
For comparison, the black histogram in each panel represents the system-to-system scatter in the $V_{\rm rot}/\sigma$ for the stellar disc.  The accreted, metal-poor, old, and/or outer spheroid subcomponents show only a mild degree of rotation and are usually dispersion-supported.  The in situ, metal-rich, young, and/or inner spheroid subcomponents show higher rotation velocities on average but there is a large system-to-system scatter in $V_{\rm rot}/\sigma$ for these subcomponents.  
}
\end{figure*}

\subsection{Rotation}

\subsubsection{Rotation - External view}

In Fig.~\ref{fig:rot_stack} we show a stacked line of sight velocity (i.e., stacked along the y-axis) contour map, where all 412 simulated galaxies have been rotated to an edge-on configuration before stacking.  As in previous plots, the star particles are interpolated to a 2D regular grid after stacking using the SPH kernel with 512 smoothing neighbours.
Red contours indicate motion away from the observer, whereas blue contours indicate motion towards the observer.  The numbers associated with the contours give the line of sight velocity in km/s.  The left panel gives a large-scale view and the right panel zooms in on the central 50 kpc.

From the colour coding of the contours alone it is evident that, on average, the spheroid has a net prograde rotation (i.e., in the same sense as the disc).  At very large radii ($r \ga 30$ kpc) the mean rotation velocity is only 10-20 km/s.  In the inner regions, however, the spheroid rotates much faster and has the appearance of an extended, `puffed up' disc that bears a close resemblance to the `vast extended stellar disc' around M31 reported by \citep{ibata05}.  The lag between the disc and the `puffed up' disc depends on the distance along the major axis but typical values range from -50 to -100 km/s, which is similar to what is found for M31.

In Fig.~\ref{fig:vrot_sigma} we show the distribution of the ratio of the rotation velocity to the 3D velocity dispersion of the spheroid of the simulated galaxies (i.e., we compute one value of $V_{\rm rot}/\sigma$ for each simulated galaxy's spheroid) for different cuts applied to the spheroid.  For comparison, in each panel the black histogram represents the system-to-system scatter in the $V_{\rm rot}/\sigma$ for the stellar disc.  Note that in computing a single value of  $V_{\rm rot}/\sigma$ for each galaxy we are averaging over all of the star particles in that galaxy (subject to the cut applied) and each particle is weighted equally in the averaging.  We are thus examining the system-to-system scatter in the {\it global} kinematics of stellar spheroids in Fig.~\ref{fig:vrot_sigma}.

In the top left panel of Fig.~\ref{fig:vrot_sigma} we compare accreted and in situ spheroid components.  Typically, the accreted component shows only a mild degree of prograde rotation (though in some cases the accreted spheroid is counter-rotating with respect to the disc) and is usually dispersion dominated ($V_{\rm rot}/\sigma$ $<<$ $1$).  The in situ spheroid, by contrast, always shows co-rotation with the disc and shows a large spread in $V_{\rm rot}/\sigma$, indicating that in some cases this component is dispersion supported and in other cases it is rotation supported (on average $V_{\rm rot}/\sigma \sim 1$).

The origin of the rotation of the in situ component is clear.  But it is interesting that the accreted component of the spheroid also typically shows some degree of prograde rotation with respect to the disc.  Why might this be the case?  Cosmological simulations have shown that there is preference for accretion of satellites along the major axis of the dark matter halo (e.g., through filaments; see \citealt{libeskind05,zentner05}).  Recently, \citet{deason11c} used the \gimic\ hydrodynamic simulations to show that there is also a general tendency (although with large scatter) for the plane of the galactic disc to lie along the major axis of the dark matter halo.  Therefore, there is some preference for satellites to be accreted parallel to the plane of the disc, which could induce a net rotation of the accreted spheroid in the same sense as the disc.  Another contributing factor may be the fact that galaxies on prograde orbits are preferentially susceptible to orbital decay due to dynamical friction (as argued by, e.g., \citealt{morrison09}), although this is only expected to be important for those satellites which are able to penetrate very close to the disc before being disrupted.  But note it has even been argued that this effect can give rise to a small net retrograde motion (e.g., \citealt{quinn86}). 

In the other panels we show the distribution of $V_{\rm rot}/\sigma$ for the spheroid component when we apply cuts in metallicity, age, or galactocentric distance.  These are meant to roughly mimic what could be done observationally to distinguish between the in situ and accreted components.  On average, the stars that are metal poor, old, or preferentially located at large radii show only a mild degree of rotation and are usually dispersion supported, as was the case for the accreted component shown in the top left panel.  Stars that are metal rich, young, or located in the inner regions of the spheroid show higher rotation velocities on average, in accordance with the in situ component shown in the top left panel.

\begin{figure}
\includegraphics[width=\columnwidth]{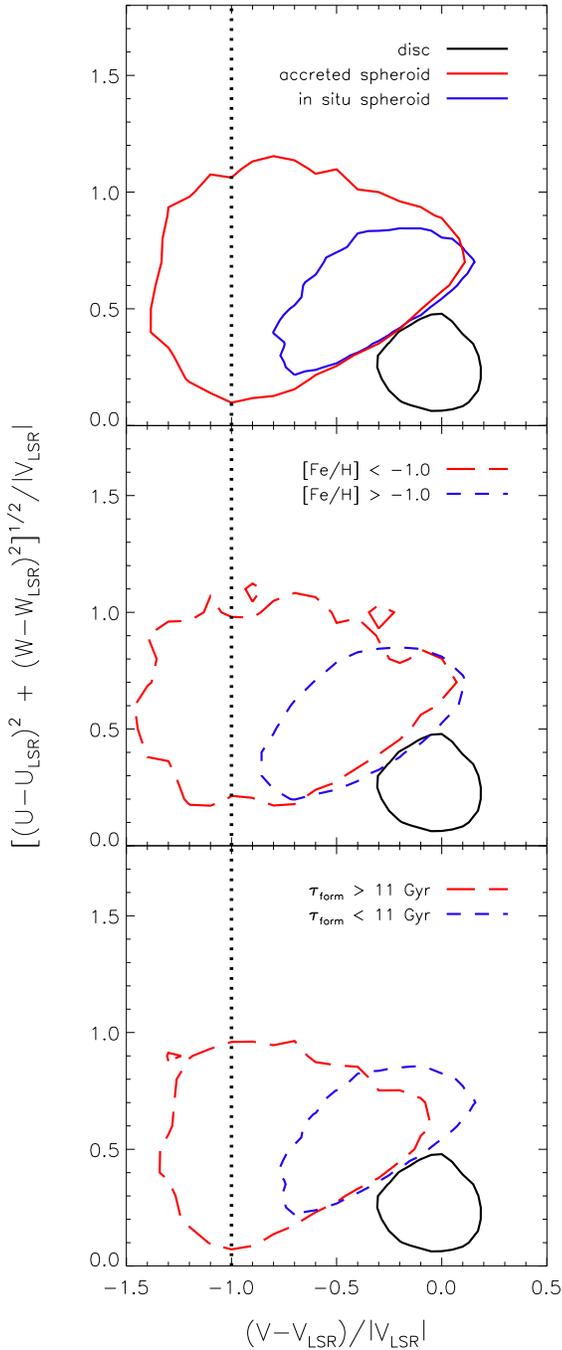}
\caption{\label{fig:toomre} 
Stacked Toomre diagram for star particles in the `Solar neighbourhood' of the simulated disc galaxies.  The contours enclose 50\% of the particles for any given subcomponent.  Left of the vertical dotted line motion is retrograde with respect to the disc. Relatively metal-rich and young stars (which are predominantly formed in situ) always rotate in prograde fashion and with a lag that ranges from 0.0 to -0.7 times the rotation velocity of the disc (see text).  Relatively metal-poor and old stars (predominantly accreted) show a much larger range of velocities with a non-negligible fraction on retrograde orbits.
}
\end{figure}

\subsubsection{Rotation - Internal view}

To facilitate comparisons with kinematic studies of the spheroid of the Milky Way, which are generally limited to the Solar neighbourhood, we show in Fig.~\ref{fig:toomre} the Toomre diagram for star particles situated within cylindrical galactocentric radii of 7.0 kpc and 10.0 kpc (the Sun is at $\approx 8.5$ kpc in our Galaxy; e.g., \citealt{vanhollebeke09,gillessen09}) and within 2 kpc of the disc midplane ($|Z| < 2$ kpc).  The velocity coordinates, $V$, $U$, and $W$ correspond to the tangential, radial, and $Z$ velocities in the rest frame of the Solar neighbourhood.  The rest frame for each of the simulated disc galaxies is computed by taking the medians of each of the three velocity components for cold star particles (i.e., those assigned to the disc) within the selected volume.  To account for the fact that our simulated galaxies span a range of disc (and total) masses, we have elected to normalise the velocities by the median rotation velocity of disc particles in the Solar neighbourhood, $V_{\rm LSR}$.  To derive the average result from the simulations, we stack the Toomre diagrams computed for each individual galaxy.  To the left of the dotted vertical line motion is retrograde with respect to the disc.

In the top panel of Fig.~\ref{fig:toomre} we compare the velocity distributions for the in situ and accreted components of the spheroid in the Solar neighbourhood.  The solid red and blue contours enclose 50\% of the particles in the in situ and accreted components, respectively.  The solid black contour encloses 50\% of the star particles in the disc component.  Since the velocity rest frame is computed from disc star particles, the disc component is situated at $V-V_{\rm LSR} \approx 0$.  The in situ component of the spheroid shows a clear prograde rotation signature with a lag ranging from -0.7 to 0.0 (median is $\approx -0.35$) times the rotation velocity of the disc.  For a disc galaxy with a rotation velocity similar to M31, $\approx 240$ km/s, the median lag would correspond to $\approx -84$ km/s.  As noted above this is similar to the velocity lag reported for the large disc-like structure around M31 by \citet{ibata05}.  The accreted component of the spheroid, by contrast, shows a much broader range of velocities, with a non-negligible fraction of stars on retrograde orbits with respect to the disc.

In the middle and bottom panels of Fig.~\ref{fig:toomre} we use metallicity and age, respectively, to effectively select the in situ and accreted components using observable quantities.  As expected, young and/or metal-rich stars have a distribution that is very similar to that of the pure in situ selection in the top panel.  Old and/or metal-poor selection yields a distribution very similar to that of the pure accreted component selection.

\begin{figure*}
\includegraphics[width=15cm]{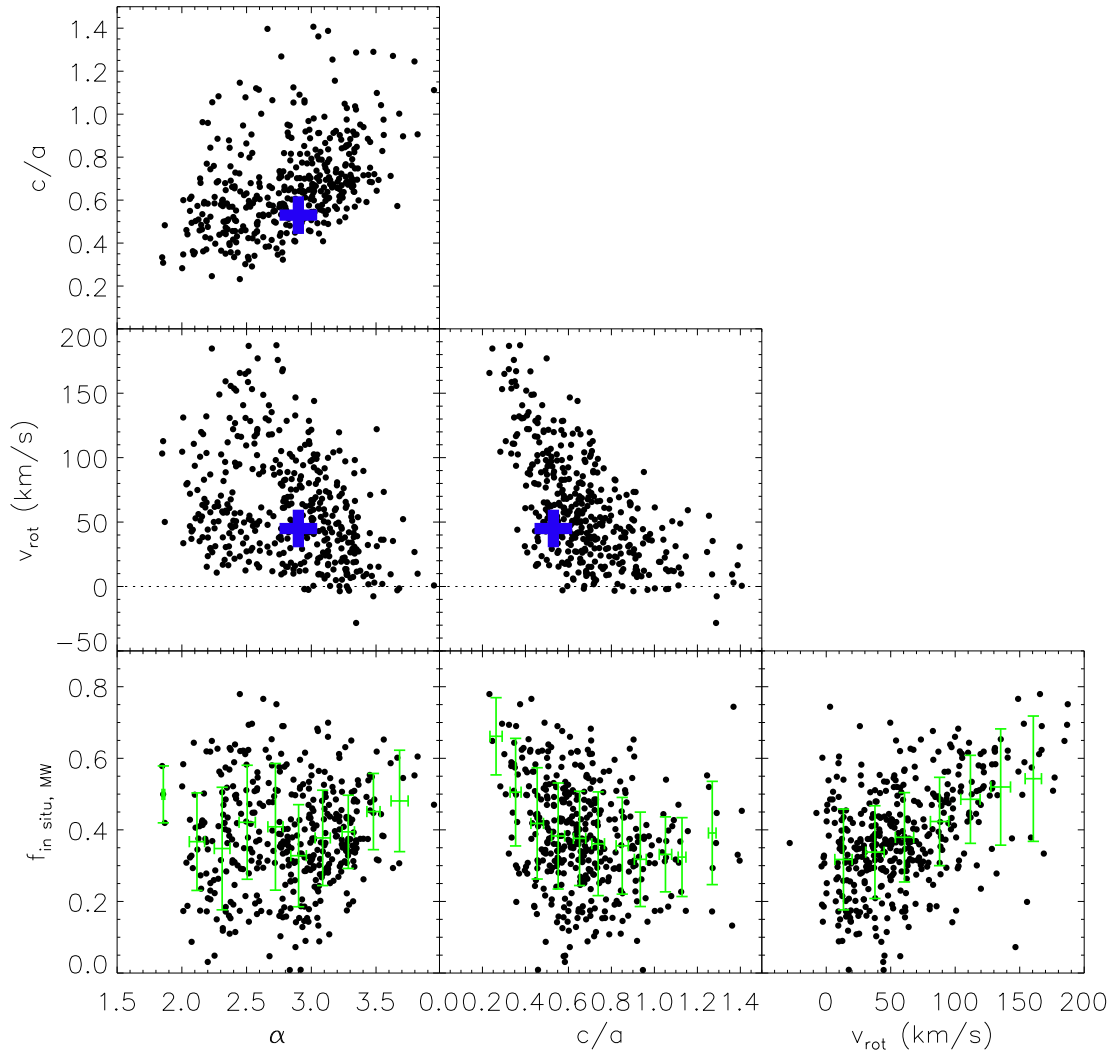}
\caption{\label{fig:shape_rot} 
Scatter plots of the best-fit shape parameters (flattening $c/a$ and power-law index $\alpha$; see Fig.~\ref{fig:shape_only}), rotation velocity ($V_{\rm rot}$), and in situ mass fraction of the spheroid (see text).
For comparison we show recent observational estimates for the shape and rotation of the Milky Way spheroid made by \citet{deason11a,deason11b}.  The top panel is the same as Fig.~\ref{fig:shape_only} but we include it for completeness.  The shape parameters, $c/a$ and $\alpha$, are clearly correlated with the rotation velocity and there is a particularly strong trend between the rotation speed and the degree of flattening of the spheroid.  The in situ mass fraction is correlated with both the flattening and the rotation velocity, in the sense that systems with a large in situ mass fraction are typically flatter and rotating faster, although there is a large degree of scatter in these trends.
}
\end{figure*}

Based on the above, we find that there is a significantly different velocity distribution for the in situ (relatively young and metal-rich) and accreted (relatively old and metal-poor) components in the `Solar neighbourhood' of the simulated disc galaxies.  Encouragingly, there is mounting observational evidence for multiple components in the kinematics of stars in our own Solar neighbourhood.  For example, \citet{carollo10} (see also \citealt{beers11}) find kinematic evidence for two components, one relatively metal-rich and one relatively metal-poor.  (We discuss these intriguing results in more detail below.)  Likewise, \citet{nissen10} (see also \citealt{nissen11}) have used a sample of 94 dwarf stars with high-resolution spectral measurements and found evidence for two components in the kinematics when they split their sample into stars with relatively high or low $[\alpha/$Fe$]$ for their [Fe/H].  They concluded that the ``high-$\alpha$'' stars origin was likely in situ.  Qualitatively our simulations are consistent with the trends found by \citet{nissen10}, although it appears that the overlap between the in situ and accreted components in Fig.~\ref{fig:toomre} is larger than that for the two components identified by \citet{nissen10} (see their Fig.\ 3).  We point out, however, that Fig.~\ref{fig:toomre} is the average (stacked) result and that we expect there to be a large range of behaviours for individual galaxies.

\subsection{Shape and Rotation together}

The shape and kinematics of a system are obviously not independent quantities, although we have treated them as such above.  We now examine the connection between the two.   

In Fig.~\ref{fig:shape_rot} we show scatter plots of the best-fit shape parameters (flattening $c/a$ and power-law index $\alpha$; see Fig.~\ref{fig:shape_only}), rotation velocity ($V_{\rm rot}$), and in situ mass fraction.
Note that the latter two quantities have been computed using the same star particle selection as in the fitting of parametric model to the mass distribution (i.e., we exclude star particles within $|Z| < 4$ kpc and beyond a 3D radius $r_{\rm 3D} > 40$ kpc).  Note also that the top panel is the same as Fig.~\ref{fig:shape_only} but we include it for completeness.  We comment on origin of the relationship between $c/a$ and $\alpha$ below.

The two panels in the middle row of Fig.~\ref{fig:shape_rot} show how $c/a$ and $\alpha$ are related to the rotation of the spheroid.  There is a strong correlation between the rotation velocity of the spheroid and its mass structure - systems with a high degree of rotation tend to be highly flattened and their mass density tends to drop off more slowly with radius than for systems with a lower degree of rotation.  The correlation between the rotation velocity and flattening is even stronger than the one between the rotation velocity and the power-law index.  Like the structural parameters, there is a large range in rotation velocities across our sample, with some spheroids rotating as fast as $150$ km/s while others are very mildly {\it counter-rotating} with respect to the disc.  The median rotation velocity of the spheroid is $+38$ km/s.

\begin{figure}
\includegraphics[width=\columnwidth]{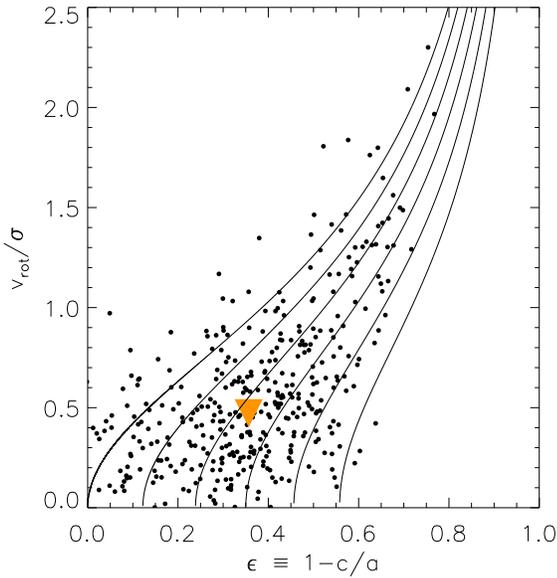}
\caption{\label{fig:binney_plot} 
Scatter plot of the ratio of the rotation velocity to the 3D velocity dispersion of the spheroid against the ellipticity of the stellar spheroid.  
The solid curves correspond to the predictions of the tensor virial theorem applied to oblate spheroids (see \citealt{binney87}, eqn. 4-95) with anisotropy parameters $\delta$ ($\equiv 1 - \sigma^2_{zz}/\sigma^2_{xx}$) = 0.0 , 0.1, 0.2, 0.3, 0.4, and 0.5 (top to bottom).  The inverted orange triangle represents the median of the simulated spheroids.  The simulated galaxies have a wide range of behaviours: for some the shape is set entirely by rotation ($\delta = 0$) while for others anisotropy plays as large a role as rotation.  The median corresponds to $\delta \approx 0.2$.  The Milky Way has $\epsilon \approx 0.3-0.5$ and $V_{\rm rot}/\sigma \la 0.5$ (see text), implying $\delta \approx 0.2-0.4$.  
}
\end{figure}

Is there a direct {\it causal} link between the rotation and the shape of the spheroid?  It is possible, for example, that velocity anisotropy is the true underlying cause of the flattened shape rather than rotation (as is the case with many giant elliptical galaxies).  We attempt to shed some light on this in Fig.~\ref{fig:binney_plot}, which shows a scatter plot of the average ratio of the rotation velocity to the 3D velocity dispersion of the spheroid against its ellipticity, $\epsilon \equiv 1 - c/a$.  The solid curves correspond to the predictions of the tensor virial theorem applied to oblate spheroids (see, e.g., \citealt{binney78,binney87}, see eqn.\ 4-95 of the latter) with anisotropy parameters $\delta$ ($\equiv 1 - \sigma^2_{zz}/\sigma^2_{xx}$) = 0.0 , 0.1, 0.2, 0.3, 0.4, and 0.5 (top to bottom).  Note that the predictions are independent of how rapidly the galaxy rotates and, if the isodensity surfaces are similar concentric ellipsoids, they are also independent of the radial density profile of the spheroid.

It is readily apparent that the spheroids of the simulated galaxies display a fairly wide range of behaviours, from being completely rotation-dominated (i.e., points clustered around the top curve) to having a significant contribution to the flattening from velocity anisotropy with $\delta$ ranging as high as $\approx 0.5$.  The median implied value of $\delta \approx 0.2$ for the simulated galaxies.  Thus, typically, both rotation and velocity anisotropy contribute to the flattening of the spheroids in our simulations.

Measuring the rotation speed of the stellar halo of the Milky Way is non-trivial as it relies on the availability of accurate distances to the spheroid tracers, an accurate distance from the Sun to the centre of the Galaxy, and finally an accurate estimate of the local standard of rest circular velocity.  While the Sun's galactocentric distance is now relatively well pinned down (to $\approx 8.5$ kpc), the other two issues remain.  \citet{carollo07,carollo10} have used a large sample of main sequence turn off stars in the SDSS to constrain the rotational velocity of both the metal-rich inner and metal-poor outer components (both confined to $r_{\rm 3D} < 40$ kpc).  Under the assumption of the canonical local standard of rest circular velocity of $220$ km/s, these authors find no evidence for rotation of the inner metal-rich component, but they find evidence for significant net {\it retrograde} motion of the outer metal-poor component of $\approx -80$ km/s.  We note that there is an ongoing debate about whether or not hard kinematic evidence exists indicating the need for at least two halo components. \citet{schoenrich11} argue that the evidence for two components found by \citet{carollo10} is spurious, and the result of errors in the distance estimates that were employed in that study.  However, as \citet{beers11} demonstrate, the stars in the sample considered by \citet{carollo10} (whether or not the unlikely  
main-sequence turn-off luminosity classifications are reassigned to dwarf and/or
giant classifications) clearly exhibit {\it proper motion distributions} that   
are asymmetric with respect to the other (non highly retrograde) stars.  \citet{beers11} conclude this illustrates that the inferred asymmetry in the highly  
retrograde tail arises not from improper distances, but rather, is due primarily
to real differences in the proper motions.

\citet{deason11a} have measured the rotation of the halo using a spectroscopic sample of BHB stars selected from the SDSS.  The advantage of BHB stars is that they are luminous and have nearly a constant absolute magnitude when selected within a certain colour range.  The disadvantages are that there are far fewer BHB stars than main sequence turn off stars to work with, and that the BHBs may be a biased tracer of the spheroid (e.g., they are typically older on average than main sequence stars).  In qualitative agreement with \citet{carollo07,carollo10}, \citet{deason11a} find evidence for a retrograde-rotating, relatively metal-poor component with $V_{\rm rot} \approx -25$ km/s.  But they also find evidence for a prograde-rotating, relatively metal-rich component with $V_{\rm rot} \approx 15 \pm 8$ km/s.  These values were derived under the assumption of the canonical local standard of rest circular velocity of $220$ km/s.  A number of recent studies (see the discussion in \citealt{deason11a}) have, however, recommended an upward revision to $\approx 240-250$ km/s.  \citet{deason11a} show that if one adopts a local standard of rest circular velocity of $240$ km/s, then the implication is that the metal-poor component has no detectable net rotation while the comparatively metal-rich component has a net prograde rotation velocity of $\approx 45$ km/s.  This is the speed that is indicated by the solid blue cross in the two panels in the middle row of Fig.~\ref{fig:shape_rot}, which is compatible with the distribution of simulated galaxies.

Is the shape of the Milky Way's stellar halo determined by rotation or velocity anisotropy?  We can use the theoretical curves plotted in Fig.~\ref{fig:binney_plot} to try to answer this question.  As discussed in Section 4.1.2, the inferred shape of the Milky Way's halo is $c/a \approx 0.5-0.7$, implying an ellipticity of $\epsilon \approx 0.3-0.5$.  Using the latest BHB-based results of \citet{deason11a} provides a measure of the rotation velocity and \citet{xue08} have estimated the line-of-sight velocity dispersion of a similar BHB sample (see their Fig.\ 10).  But note that what is plotted in Fig.~\ref{fig:binney_plot} is the ratio of the rotation velocity to the {\it 3D} velocity dispersion, which is not directly observable.  Thus, one must assume something about the degree of anisotropy in order to infer an anisotropy from Fig.~\ref{fig:binney_plot}, which is clearly not self-consistent.  However, it can be said with some certainty from the BHB results that $V_{\rm rot}/\sigma \la 0.5$ for the Milky Way's stellar halo.  Using the theoretical curves in Fig.~\ref{fig:binney_plot}, implies $\delta \approx 0.2-0.4$, which is consistent with the implied spread in the spheroids of the simulated galaxies.  Thus, based on recent BHB measurements (which show the halo has a non-zero rotation speed), one infers that both rotation and velocity anisotropy contribute to the flattening of the Milky Way's stellar halo.  

We can now return to the top panel of Fig.~\ref{fig:shape_rot}, to try to explain the relationship between $c/a$ and $\alpha$ in terms of the correlations of these parameters with the rotational speed, $V_{\rm rot}$.  There is a clear lower envelope to the distribution in the top panel: the minimum $c/a$ increases with $\alpha$ (or, equivalently, the maximum $\alpha$ increases with $c/a$).  This same envelope is also visible in the middle left panel: the maximum $\alpha$ increases with decreasing $V_{\rm rot}$ and the central panel shows that $c/a$ increases with decreasing $V_{\rm rot}$. Thus, the maximum $\alpha$ must increase with increasing $c/a$.  However, this kind of correlation argument does not really explain the trend between $c/a$ and $\alpha$ or demonstrate causality.  A possible explanation may be found in the $V_{\rm rot} - \alpha$ plane.  In particular, for a given $\alpha$, $V_{\rm rot}$ cannot be arbitrarily high. If it is too high, the centripetal force would exceed the gravitational one (which depends on $\alpha$). 

Finally, in the bottom row of Fig.~\ref{fig:shape_rot} we investigate whether the shape of the stellar mass distribution or its kinematics tell us about the in situ mass fraction.  The green points and error bars represent the median and 1-sigma (i.e., encloses 68\% of the data) in the in situ mass fraction in bins of $\alpha$, $c/a$, and $V_{\rm rot}$.  As noted above, the in situ mass fraction has been computed within the volume limits specified above.  

There is a wide system-to-system variation in the in situ mass fraction, with typical fractions in the range of 20-60\% (median of $\approx$ 40\%).  We find no strong trend between the in situ mass fraction and the power-law index, $\alpha$.  A trend exists between the in situ mass fraction and the flattening of the spheroid for $c/a \la 0.6$ or so, in that systems with large in situ mass fractions are typically flatter than those with low in situ mass fractions.  An even stronger trend is evident between the in situ mass fraction and the rotation velocity of the stellar spheroid.  If one uses the simulations as a guide to infer an in situ mass fraction for the Milky Way, the shape and kinematics would be consistent with a mass fraction of $\sim$ 40\%.  

\section{Summary}

We have used the \gimic\ simulations to study the global structure and kinematics of the spheroids of $412$ simulated $\sim L*$ disc galaxies.  \citet{font11} have recently demonstrated that these simulations are able to successfully reproduce the metallicity and surface brightness profiles of the spheroids of the Milky Way and M31.  A key to the success of the simulations is a significant contribution to the spheroid from stars that formed {\it in situ}.  While the outer halo is dominated by accreted stars, stars formed in the main progenitor of the galaxy dominate at $r \la 30$ kpc.  

In the present study we have shown that there is a close physical connection between the in situ spheroid and the disc, namely that the in situ spheroid was produced by the dynamical heating of a proto-disc.  The likely mechanism for the heating is accretion of satellites (and mass in general) at $z \sim 1$.  A similar conclusion was reached recently by \citet{purcell10} on the basis of the analysis of idealised simulations of disc bombardment by infalling satellites.

Given the very different origins for the in situ and accreted spheroids, we expect there to be differences in the global structure and kinematics of the two components.  Indeed, we have shown that the inner regions of the spheroid are highly flattened (oblate) and show evidence for two components in the kinematics, with the dominant one rotating in prograde fashion with respect to the disc.  The degree of flattening is in good accordance with detailed observations of both local galaxies, such as the Milky Way (e.g., \citealt{chiba01,juric08,deason11b,sesar11}) and M31 \citep{pritchet94,ibata05}, and stacked observations of more distant galaxies \citep{zibetti04}.  This is in stark contrast to the results of collisionless hybrid models in which star particles are `painted on' to dark matter haloes of infalling dwarf galaxies (e.g., \citealt{bullock05,font06a,delucia08,cooper10}), which usually give rise to highly {\it prolate} configurations, like that of the underlying dark matter halo \citep{cooper10}.  

As we noted in the Introduction, hybrid models also fail to reproduce the observed radial trend with metallicity found in the Milky Way and M31 \citep{font06b,delucia08,cooper10}, and it is difficult to see how they could be easily reconciled with the kinematic and chemodynamic evidence for a `dual halo' in the Milky Way (e.g., \citealt{carollo07,carollo10,nissen10}).  The \gimic\ simulations, on the other hand, successfully and simultaneously account for the large-scale metallicity distribution, the flattening of the spheroid, and the presence of two components in the kinematics of the spheroid stars.  

Including hydrodynamics in the simulations does not in and of itself guarantee success though.  We suggest that the stars that constitute the present-day in situ spheroid must be formed in a proto-disc (at relatively late times of $z \sim 1-1.5$) in order for the present-day spheroid to have the correct shape and kinematics.  This likely requires an efficient form of feedback to operate in order to delay the formation of these stars.  This is achieved in the \gimic\ simulations with efficient, but energetically feasible, supernova feedback.  Note that strong SN feedback is required in simulations in order to explain the generally low star formation efficiency and the `missing baryon problem' of normal galaxies, and to enrich the intergalactic medium via outflows. 

As a further test, it will be interesting to see whether such hydrodynamic simulations can also account for the relatively low fraction of unbound substructure (relative to that predicted by hybrid models) that is observed in the Milky Way (e.g., \citealt{xue11}).  We intend to explore this in a future study.

\section*{Acknowledgements}

The authors thank the anonymous referee for a constructive report that improved the paper. 
IGM is supported by a Kavli Institute Fellowship at the University of Cambridge.
ASF is supported by a Royal Society Dorothy Hodgkin Fellowship at the University of Cambridge.  RAC is supported by the Australian Research Council via a Discovery Project grant. AJD thanks the Science and Technology Facilities Council (STFC) for the award of a studentship.  The simulations presented here were carried out by the Virgo Consortium for Cosmological Supercomputer Simulations using the HPCx facility at the Edinburgh Parallel Computing Centre (EPCC) as part of the EC's DEISA `Extreme Computing Initiative', the Cosmology Machine at the Institute for Computational Cosmology of Durham University, and on Darwin at the University of Cambridge.  This work was supported by Marie Curie Initial training Network CosmoComp (PITN-GA-2009-238536).

\section*{Appendix: Numerical convergence}

Here we investigate the sensitivity of our results to numerical resolution.  As noted in Section 2, the $-2\sigma$ sphere has been simulated at eight times better mass resolution than the intermediate-resolution runs analysed in the present study.  

\begin{figure*}
\includegraphics[width=\columnwidth]{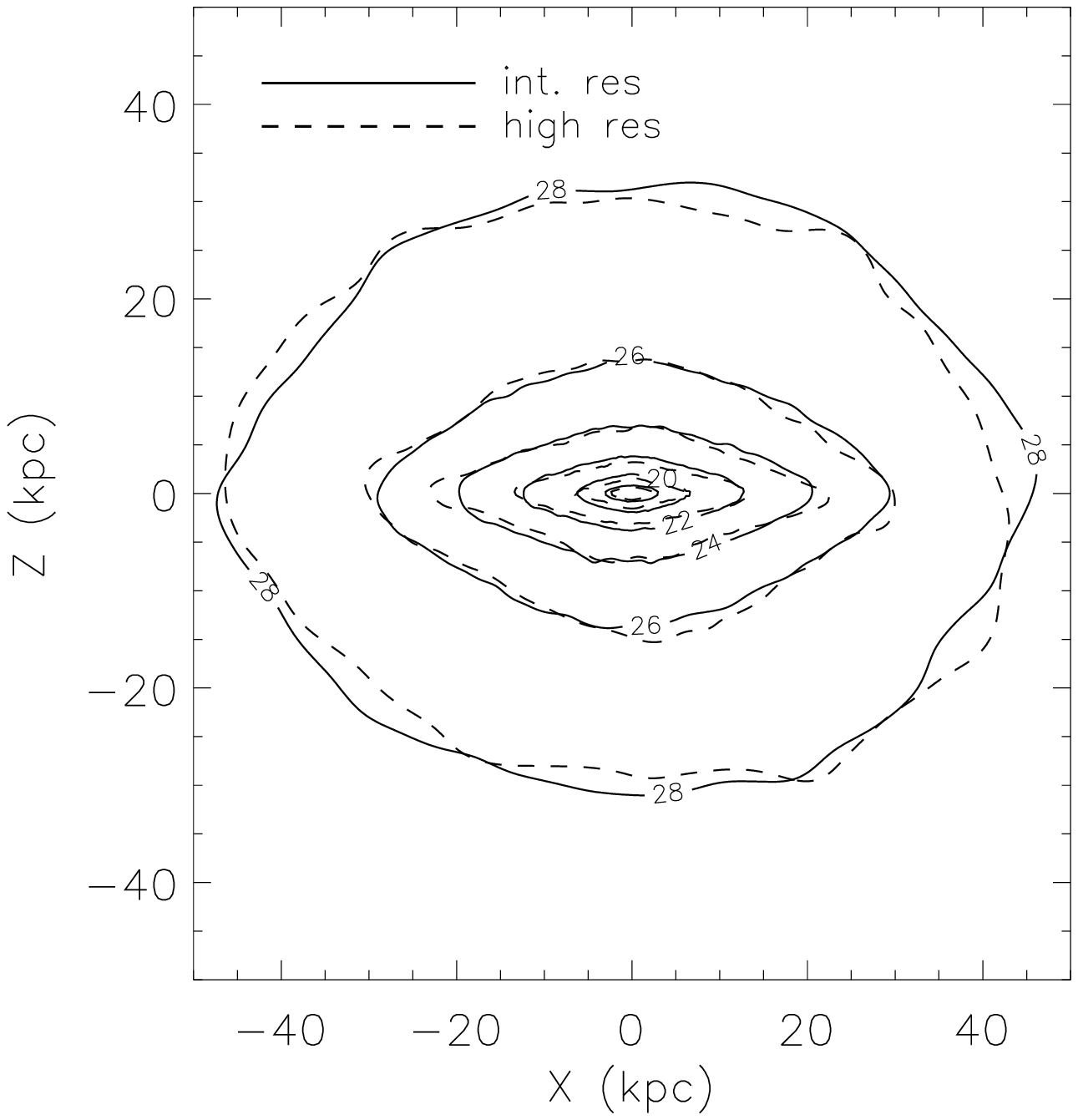}
\includegraphics[width=\columnwidth]{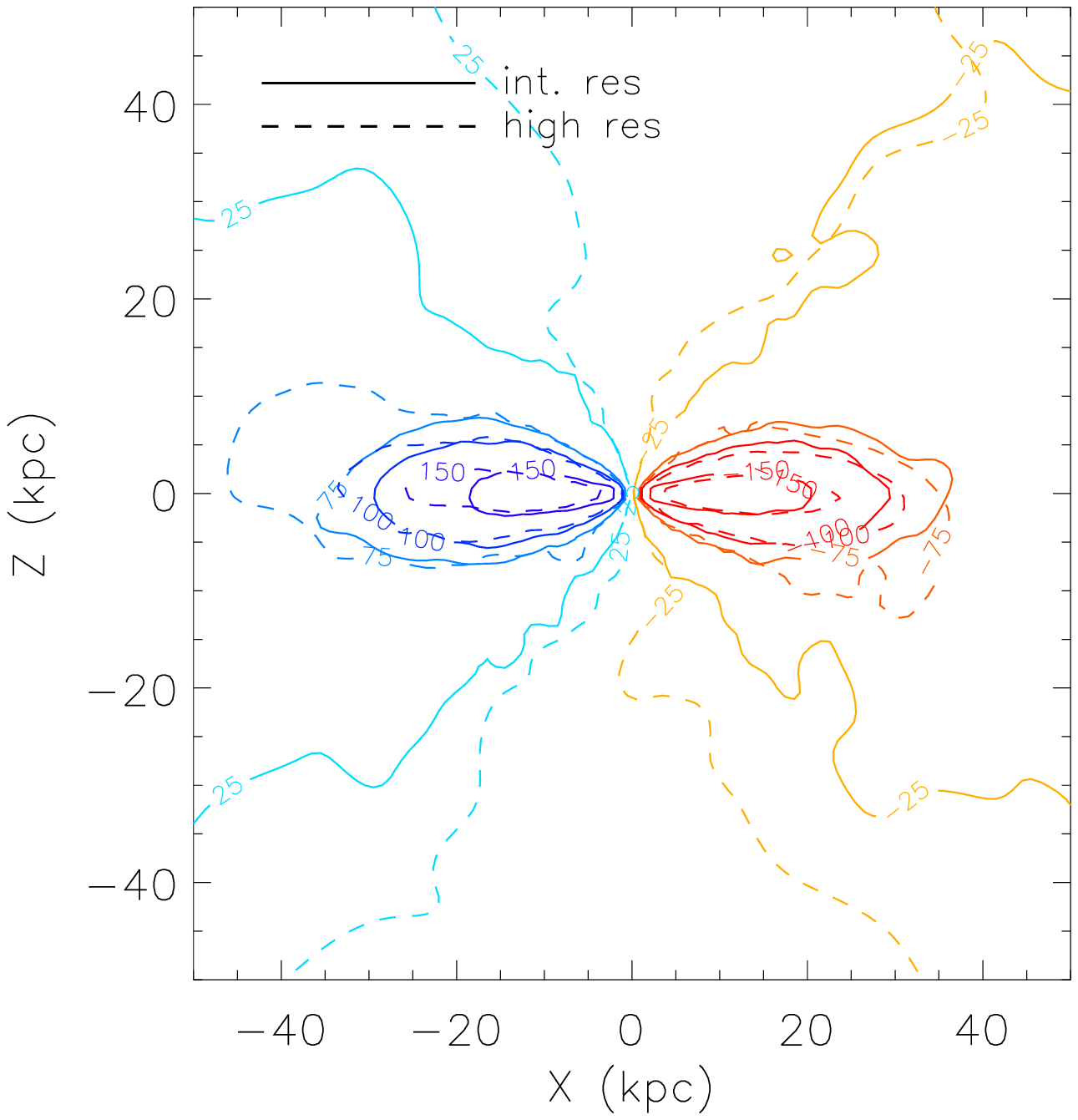}
\caption{\label{fig:res}
{\it Left:} Stacked SDSS i-band surface brightness contour maps of the simulated galaxies, as in Fig.~\ref{fig:shape_iband}, for the intermediate-resolution (solid curves) and high-resolution (dashed curves) simulations. 
{\it Right:} Stacked line of sight ($v_y$) velocity contour maps, as in Fig.~\ref{fig:rot_stack} for the intermediate-resolution (solid curves) and high-resolution (dashed curves) simulations.  Bluer colours indicate motion towards the observer and redder colours indicate motion away from the observer.  Overall there is very good agreement between the mean structure and kinematics of the intermediate- and high-resolution galaxy samples.
}
\end{figure*}

In Fig.~\ref{fig:res} we compare the mean structure and kinematics of 50 Milky Way-mass disc galaxies in the high-resolution simulation with those of the 412 galaxies in the intermediate-resolution runs.  
Overall, the agreement between the high-resolution and intermediate-resolution \gimic\ simulations is quite good for both mean structure and kinematics.

\end{document}